\pdfoutput=1

\documentclass[aps,pra,reprint,showpacs]{revtex4-1}
\usepackage{graphicx}
\usepackage{amsmath}
\usepackage{amssymb}
\usepackage{amsfonts}
\usepackage{latexsym}
\usepackage{rotating}
\usepackage{xcolor}
\usepackage{latexsym}
\usepackage{epsfig}
\usepackage{epstopdf}
\usepackage{natbib}
\usepackage{bm}
\usepackage{url}
\usepackage{lineno}

\usepackage{soul}

\usepackage{color} 

\usepackage{hyperref}
\hypersetup{
colorlinks=true,final=true,
        linkcolor=blue,
        citecolor=blue,
        filecolor=blue,
        urlcolor=blue,
}

\begin{document}

\title{Stability and variational analysis of cavity solitons under various perturbations}
\author{Ambaresh Sahoo}
\email{ambaresh@phy.iitkgp.ernet.in}
\author{Samudra Roy}
\email{samudra.roy@phy.iitkgp.ac.in}
\affiliation{Department of Physics, Indian Institute of Technology Kharagpur, West Bengal 721302, India}

\begin{abstract}
We theoretically investigate the dynamics and stability of a temporal cavity soliton (CS) excited inside a silicon-based microresonator that exhibits free-carrier generation as a result of two-photon absorption (TPA). The optical propagation of the CS is modeled through a mean-field Lugiato-Lefever equation (LLE) coupled with an ordinary differential equation accounting for the generation of free carriers owing to TPA. The CS experiences several perturbations (like intrapulse Raman scattering (IRS), TPA, free-carrier absorption (FCA), free-carrier dispersion (FCD), etc.) during its round-trip evolution inside the cavity. We develop a full variational analysis based on a Ritz optimization principle which is useful in deriving simple analytical expressions describing the dynamics of individual pulse parameters of the CS under perturbation. TPA and FCA limit the efficient comb generation and modify the stability condition of the CS. We determine the critical condition of stability modified due to TPA and derive closed-form expressions of the saturated amplitude and width of stable CS. We perform detailed modulation-instability analysis and obtain  stability condition against perturbations of steady-state solution of LLE. The CS experiences FCD which leads to a temporal acceleration resulting in spectral blueshift. Exploiting the variational analysis, we estimate these temporal and spectral shifts. We also include IRS in our perturbation theory and analytically estimate the frequency redshifting. Finally, we study the effect of pump-phase-modulation on a stable CS. All our analytical results are found to be in good agreement with the data obtained from the full numerical solution of LLE.

\end{abstract}

\maketitle

\section{Introduction}
\noindent Cavity solitons (CSs) are special types of dissipative solitons that persist in  driven passive cavity systems, such as fiber ring cavities and monolithic microresonators, and are ideal candidates for optical frequency comb generation and all-optical buffers applications \cite{grelu,Leo}. Being a soliton, such stable structures do not spread or dissipate. The group velocity dispersion (GVD) is balanced by nonlinearity whereas, the losses are compensated by an external driving field \cite{coen,chembo,coen_dw}. The dynamics of a CS is governed by the mean-field Lugiato-Lefever equation (LLE) \cite{LL} which generally doesn't contain any higher-order dispersion or nonlinear terms.  However, the important effect that comes into the picture for ultrashort pulses is the intrapulse Raman scattering (IRS) which causes the self-frequency redshift \cite{Vahala,Kippenberg, Wabnitz, Erkintalo, CM-DV}. Hence in realistic systems, the CSs are not only influenced by higher-order GVD \cite{coen_dw, Milian}, but also by higher-order nonlinear effects in terms of IRS and self-steepening. Owing to its large nonlinearity and wide transparency, Silicon (Si) is highly advantageous as a photonic platform for integrated optical devices like microresonator \cite{Griffith}. However, for Si-based microresonator, the presence of two-photon absorption (TPA) is relevant in the wavelength range 0.8$\,\mu$m$\,<\lambda_0<\,$2.2$\,\mu$m \cite{Hansson,Yin}, which causes free-carrier (FC) generation. Free carriers limit the efficiency of the device by introducing free-carrier absorption (FCA) and also change the refractive index through free-carrier dispersion (FCD) that results in a nonlinear cavity detuning \cite{Hansson,Lau}. The stability of the CS is sensitive to operating parameters (like detuning, CW pump strength, etc.) and modified under perturbations. The perturbations mentioned here are mainly device-oriented and we do not have much control over them.  In order to investigate the role of various perturbations on a CS, we introduce a semianalytical treatment based on Lagrangian analysis. Further, we extend our analysis to understand the dynamics of CS when the driving field is phase-modulated \cite{JKJ-SC} and find that no CS will exist beyond a critical value of the phase of the driving pump. We address each perturbation  separately to study the stability  of a perturbed CS persisting inside a nonlinear resonator. The study is based on the mean-field steady-state CW bistability analysis \cite{gibbs} that leads to different regions for the intracavity field \cite{coen}. Exploiting the CW bistability analysis we derive the modified stability condition of the system parameters for the existence of the CS under the influence of TPA, FCA, and IRS. We further develop the modulation-instability (MI) analysis \cite{Haelterman,TH-DM,Hamerly,Halder,barashenkov96,Parra-Rivas18} for all perturbations and obtain the range of values of the system parameters that define the stability conditions against perturbation of the steady-state solution of LLE, and also obtain stable and unstable regions of the bistability curve. The dynamics of perturbed CS is theoretically investigated by adopting Lagrange's variational technique \cite{Bondeson,Anderson,Kaup,Cerda,Royjlt,Cardoso2017,Cardoso_scirep2017} based on a Ritz optimization, where we assume an initial pulse shape having different parameters (i.e. amplitude, width, phase, etc.) that evolves over a round-trip time. This analytical process leads to equations of motion of different pulse parameters in the form of coupled ordinary differential equations (ODEs). By decoupling the ODEs using suitable approximations, we deduce closed-form expressions of the pulse parameters revealing interesting physics.  The variational technique is a standard analytical method used both in conservative and dissipative systems.  A Lagrangian density is introduced that includes the conservative parts where the dissipative terms are considered as perturbations. We confirm that the physical values of all perturbations considered here are small and valid under perturbative analysis. The small perturbation ensures the nominal structural deformations of the propagating pulse. The variational method relies on the proper choice of the ansatz which retains its shape during propagation. The ansatz is well-defined for Kerr solitons governed by a nonlinear Schr\"{o}dinger equation (NLSE) \cite{GPA}. We can choose a Pereira-Stenflo type solution as an ansatz for dissipative systems which is governed by complex Ginzburg-Landau equation (GLE) \cite{AS}. For a nonlinear resonator, however, we do not have an exact mathematical expression of a temporal CS that forms over a CW background. Hence, we rely on the $sech$ function as our trial ansatz for variational treatment which can predict the typical characteristics of a CS under perturbations with good accuracy and reveals interesting physics.

\section{Mean-field model}
\label{Ginzburg--Landau Equation}
\noindent The nonlinear passive cavity dynamics in the presence of higher-order nonlinear effects with intracavity field amplitude $u(t,\tau)$ is modeled through a mean-field normalised LLE (a damped-driven NLSE) \cite{LL,Vahala,KEW,Kippenberg,Hansson,Erkintalo,GPA} which includes higher-order dispersion, self-steepening, IRS, TPA and FC effects as
\begin{align} \label{LL}
\frac{\partial u}{\partial t }=  \left[-1 - i\,\Delta + i\sum_{n\geq2}{\delta_n \left(i \frac{\partial}{\partial\tau} \right)^n}  \right]u +S \nonumber\\ 
+\,i \left[\left(1+iK\right)|u|^2 u + i\tau_{sh}\frac{\partial (|u|^2 u) }{\partial \tau} -\tau_R u \frac{\partial |u|^2 }{\partial \tau} \right. \nonumber \\ \left. +\left(\frac{i}{2} -\mu \right)\phi_c u \right], 
\end{align}
where the FC effects are included through the rate equation for the normalized carrier density $\phi_c$ \cite{Lin}
\begin{equation} \label{ansatz}
   \frac{d\phi_c}{d\tau} =\theta|u|^4-\tau_c \phi_c,
\end{equation}
with the following boundary condition $\phi_c(t,-\mathcal{T_R}/2)=\phi_c(t+\Delta t,+\mathcal{T_R}/2)$ (i.e., cumulative accumulation of FC density is considered over the round trips) \cite{Lau}. The parameters used in Eq.\,\eqref{LL} are rescaled and given in Table-I. In the absence of TPA ($K=0$), higher-order dispersion ($\delta_{n>2}=0$), IRS ($\tau_R=0$), self-steepening ($\tau_{sh}=0$) and free carriers ($\theta=0$), Eq.\,\eqref{LL} reduces to the standard LLE, whose temporally localized steady-state solution admits \textit{unperturbed CS} \cite{coen, MA-SGM}.
Here, we consider the GVD to be negative ($\delta_2=-1$) and neglect all the higher-order dispersion terms ($\delta_{n>2}=0$) for simplicity. The standard split-step Fourier method \cite{GPA} is used to solve this inhomogeneous NLSE numerically by launching a standard $sech$ pulse in the anomalous dispersion domain. Note that, unlike conservative Kerr solitons (governed by NLSE) \cite{GPA} or Pereira-Stenflo type dissipative solitons (governed by GLE) \cite{AS}, the CS does not have any well-defined mathematical structure hence we rely on the $sech$ pulse shape as an input for numerical simulation.\\ 
\begin{table}[h!]
  \begin{center}
   \caption{The rescaled parameters of Eq.\,\eqref{LL}}
   \label{tab:table1}
   \begin{tabular}{l c r} 
   \hline
   \textbf{Description}& \textbf{\hspace{-1.5cm}Rescaled/normalized as} & \textbf{Refs.}\\      
   \hline
   Slow time, $t$ &\hspace{-1.5cm} $\alpha t/t_R \rightarrow t$ & \cite{coen} \\
   Fast time, $\tau$ &\hspace{-1.5cm} $\tau \sqrt{2\alpha/(|\beta_2(\omega_0)|L)}\rightarrow \tau$ &,,\\
   Fast time normalization\\ time scale, $\tau_s$ & \hspace{-1.5cm}$\tau_s= \sqrt{|\beta_2(\omega_0)|L/(2\alpha)}$ & ,,\\
   Intracavity field \\amplitude, $A$ &\hspace{-1.5cm} $u= A\sqrt{\gamma_R L/\alpha}$ &,,\\
   Driving field \\strength, $A_{in}$ &\hspace{-1.5cm} $S = A_{in}\sqrt{\gamma_R L \Theta/\alpha^3}$ & ,,\\
   Phase detuning, $\delta_0$  &\hspace{-1.5cm} $\Delta = \delta_0/\alpha$ & ,,\\
   $n^{th}$ order dispersion\\ parameter,  $\beta_n(\omega)$ & \hspace{-1.0cm}$\delta_n=2\beta_n(\omega)/\left(n!|\beta_2(\omega_0)|\tau_s^{n-2}\right)$ & ,,\\
   Round-trip time, $t_R$  &\hspace{-1.5cm} $\mathcal{T_R}=t_R/\tau_s$ & ,,\\
   IRS parameter, $T_R$ & \hspace{-1.5cm}$\tau_R=T_R/\tau_s$ & \cite{GPA}\\
   Self-steepening parameter & \hspace{-1.5cm}$\tau_{sh}=1/(\omega_0 \tau_s)$ & ,,\\
   TPA coefficient & \hspace{-1.0cm}$K=\gamma_I/\gamma_R=\beta_{TPA}\lambda_0/(4\pi n_2)$ & \cite{Yin}\\
   FC density, $N_c$  &\hspace{-1.5cm}  $\phi_c=\mathcal{C} N_c L/\alpha$ & \cite{Lau}\\
   FC generation term   &\hspace{-0.9cm} $\theta=\beta_{_{TPA}} \mathcal{C} \tau_s \alpha/(2\hbar\omega_0 A_{_{eff}}^2 L \gamma_R^2)$ & ,,\\
   FCD coefficient &\hspace{-1.5cm} $\mu=2\pi k_c/(\mathcal{C} \lambda_0)$ & ,,\\
   FC recombination time, $t_c$  & \hspace{-1.5cm}$\tau_c=\tau_s/t_c$ & \\
 \hline
\end{tabular}
\end{center}
\end{table}\\
\begin{table}[h!]
 \begin{tabular}{p{8.5cm}}
\vspace{-1.75cm} 
$L$, $\alpha$ and $\Theta$ are the cavity round-trip length, total cavity loss and coupling power transmission coefficient, respectively. The silicon nonlinear parameter $\gamma=\gamma_R +i\gamma_I$, with $\gamma_R=2\pi n_2/\left(\lambda_0 \,A_{eff}\right)$ and $\gamma_I=\beta_{TPA}/(2A_{eff})$, where $n_2\approx(4\pm 1.5)\times 10^{-18} \ m^2 W^{-1}$ and $\beta_{TPA}\approx 8\times 10^{-12} \ m W^{-1}$.
At $\lambda_0 (=2\pi c/\omega_0)=1.55~\mu m$, the FCA cross section is $\mathcal{C}\approx 1.45\times 10^{-21}\ m^2$ and  $k_c\approx 1.35\times 10^{-27}\ m^3$ \cite{Dinu,Rong,Lau}.\\
\hline
  \end{tabular}
\end{table}\\

\section{VARIATIONAL ANALYSIS}
\label{perturbative}
\noindent The governing LLE [Eq.\,\eqref{LL}] contains TPA, free carriers and higher-order nonlinear terms as perturbations. The important question is how these terms affect the stability and dynamics of the CS solution of LLE. One can study their impact by simply solving Eq.\ \eqref{LL} numerically, but this approach unveil limited physical insight. We treat the four terms as small perturbations and study their impact theoretically through a variational analysis. The variational method has been used with success in the past for many pulse-propagation problems \cite{Bondeson,Kaup,Anderson,Cerda,Royjlt,Cardoso2017,Cardoso_scirep2017} where a suitable \textit{ansatz} for the pulse shape is required. The technique is based on the assumption that the functional form of the pulse shape remains intact in presence of small perturbations but its parameters appearing in the ansatz (amplitude, width, position, phase, frequency, etc.) may vary with propagation. In our case, the perturbation theory is developed by introducing the \textit{ansatz} \cite{Vahala,Wabnitz_ansatz},
\begin{align}\label{nls0}
 u\left( t,\tau  \right)= \sqrt{\frac{E(t)\eta(t)}{2}}  \text{sech}\left\{ \eta \left( t  \right)\left[ \tau -{{\tau }_{p}}\left( t  \right) \right] \right\} \nonumber \\
\times \exp\left\{ i \phi \left( t  \right)-i\Omega_p\left( t  \right)\left[ \tau -{{\tau }_{p}}\left( t \right) \right]   \right\},
\end{align}
where the five parameters  $E$ (pulse energy), $\eta$ (inverse of temporal pulse width), $\tau_p$ (temporal peak position), $\phi$ (phase) and $\Omega_p$ (frequency peak position) are now assumed to evolve with slow time $t$.  The actual CSs are the localized pulses sitting on top of a CW background.  The ansatz we consider here, however, does not include any chirp and background. We compromise the chirp and background term because in the presence of these two parameters, Eq.\,\eqref{nls4} becomes non-integrable and it is difficult to deduce any closed-form expressions which we aim for. However, the present form of the ansatz allows for the temporal and spectral shifts of CSs.
The variational method results in the following set of five coupled (four ordinary differential equations (ODEs) and one self-consistent equation) equations (see APPENDIX):
\begin{align} 
\frac{dE}{dt}&= -2E \nonumber  - \frac{2}{3}K\eta E^2 -\frac{1}{6}\theta\eta E^3 +2S\left(\frac{E}{2\eta}\right)^{1/2} \pi \\ &\hspace{3.5cm}
\times {\rm sech}{\left(\frac{\pi \Omega_p}{2 \eta}  \right)}\,\cos{\phi}, \label{var6} \\
 \frac{d{{\tau}_{p}}}{dt }& = -2\Omega_p +\frac{1}{2}\tau_{sh} E\eta-\frac{7}{72}\theta E^2, \label{var7}\\
 \frac{d{{\Omega_p }}}{dt }&=-\frac{4}{15}\tau_R E \eta^3 +\frac{2}{15}\mu \theta (E\eta)^2 -2S\left(\frac{\eta}{2E}\right)^{1/2} \frac{\pi \Omega_p}{\eta}   \nonumber\\
 & \hspace{3.5cm}\times {\rm sech}{\left(\frac{\pi \Omega_p}{2 \eta}  \right)}\,\cos{\phi}, \label{var8} \\
\frac{d\phi}{dt} &= - S\left(\frac{1}{2E\eta}\right)^{1/2} \pi \, {\rm sech}{\left(\frac{\pi \Omega_p}{2 \eta}  \right)}\,\sin{\phi}  -\Delta + \Omega_p^2 \nonumber \\ &\hspace{-0.0cm} + \frac{1}{3}\eta\left(E-\eta \right)-\frac{1}{6}\tau_{sh}E\eta\Omega_p +\frac{1}{6}\theta E^2\left( \frac{7}{12}\Omega_p -\mu \eta \right), \label{var9}\\
\eta = &\frac{E}{4}\left(1+\tau_{sh}\Omega_p + \frac{\mu \theta}{2} E \right) +\frac{3}{2\eta}S\left(\frac{1}{2E\eta}\right)^{1/2}
 \nonumber \\
 & \hspace{-0.3cm} \times \pi \, {\rm sech}\left(\frac{\pi \Omega_p}{2 \eta}  \right)\left[1- \frac{\pi \Omega_p}{\eta}{\rm tanh}{\left(\frac{\pi \Omega_p}{2 \eta}  \right)} \right] \sin{\phi}. \label{var10}
\end{align}
These equations provide considerable physical insight since they show which perturbations affect a specific pulse parameter. For example, the Raman parameter $\tau_R$ and FC parameter $\theta$ appear in the equation for the frequency shift $\Omega_p$ and the terms containing them have opposite signs. This immediately shows that the IRS leads to a spectral redshift of the CS, whereas the FC effect counterbalances this by imposing a blueshift. The total energy $E$ [Eq.\,\eqref{var6}] of the CS is affected by linear, TPA, and FCA losses which is compensated by the driving field ($S$) and eventually forms a steady-state where $dE/dt=0$. Exploiting the variational result we find that before achieving a steady value, $E$ experiences a damped oscillation. The temporal shift of the CS is also affected by the free carriers and Raman scattering. These kinds of physical insights are valuable in interpreting numerical results. In the following sections, we investigate the effects of individual perturbation on the stability and dynamics of the CS. We also compare the analytical results based on variational treatment with the full numerical simulations of Eq.\ \eqref{LL}.

\section{Impact of two-photon absorption on cavity soliton}
In semiconductor-based microresonators, the presence of TPA is relevant which modifies the stability criteria of the system and leads to a change in amplitude and pulse width of the CS. Using stability analysis based on the CW bistability analysis \cite{gibbs} and MI analysis \cite{Haelterman} we find that in the presence of TPA the threshold values of the parameters are modified and as a result the dynamics of CS is also influenced by them. We also find that, for a given set of input parameters, there exists an upper limit of the TPA coefficient ($K_{\rm max}$) beyond which CS ceases to exist. 
\begin{figure}[b!]
\begin{center}
\epsfig{file=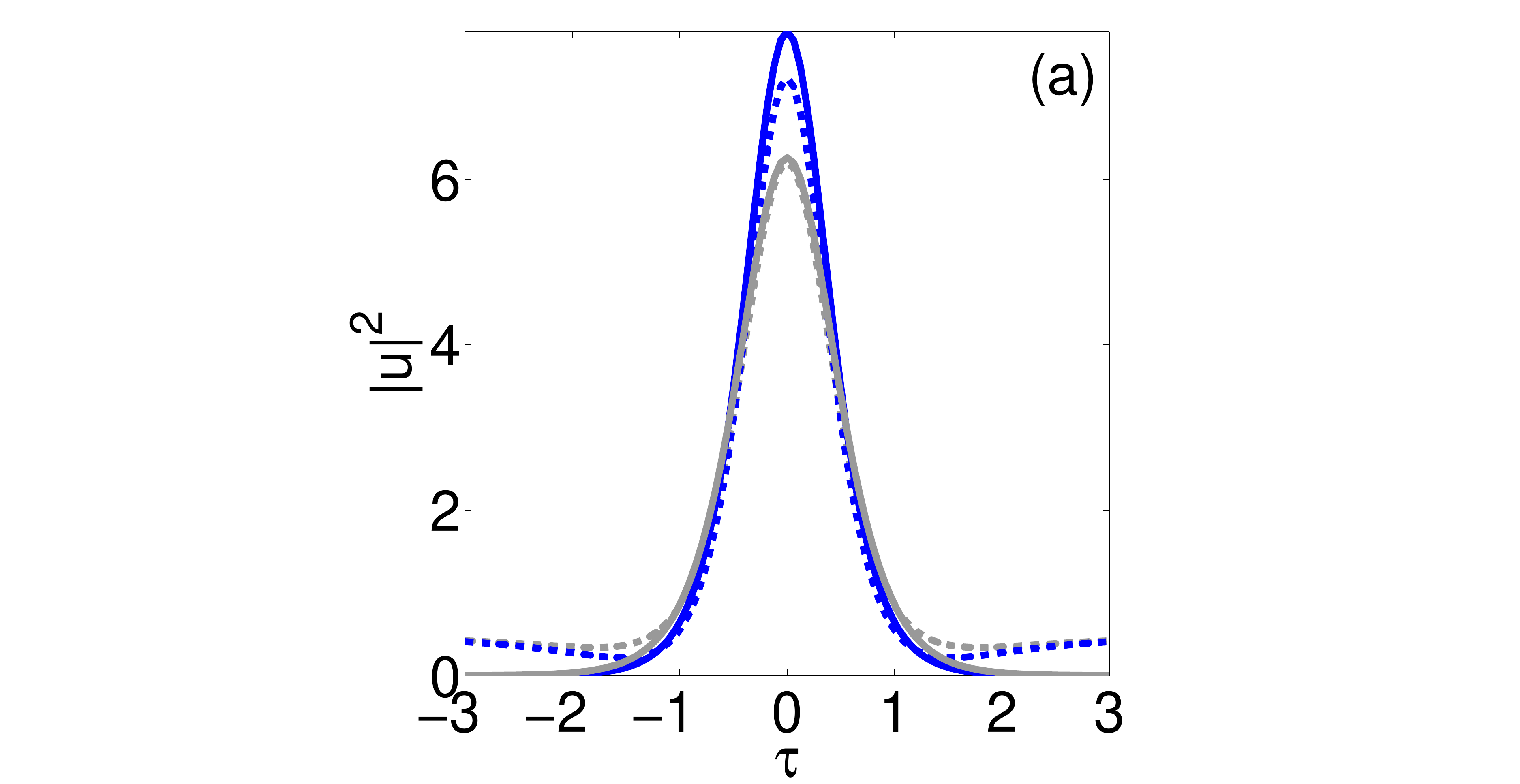,trim=0.0in 0.0in 0.0in 0.0in,clip=true, width=38mm}
\epsfig{file=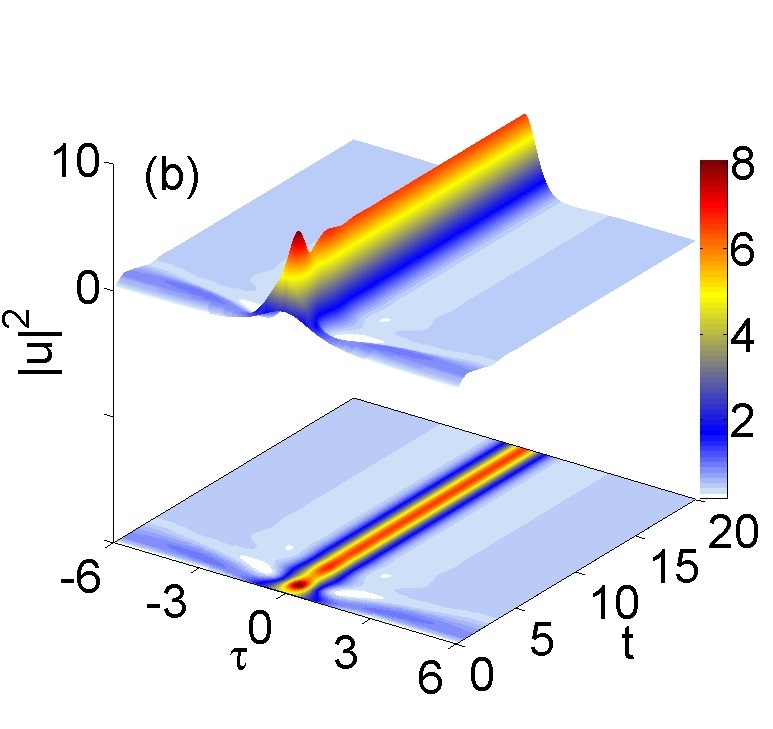,trim=0.0in 0.0in 0.0in 0in,clip=true, width=47.0mm}
\caption{(Color online) (a) Formation of CS in presence ($K=0.04$) and in absence ($K=0$) of TPA  at  $t=20$ is represented by dashed-gray and dashed-blue lines respectively, when $sech$ pulse is used as input. The solid gray and blue lines give the variational predictions of the pulse shape for ($K=0.04$) and ($K=0$) respectively. (b) Temporal evolution of CS over time $t$ for $K=0.03$ with $\Delta=3;~S=\sqrt{3.5}$.}\label{figTPAevolution}
\end{center}
\end{figure}
%
In Fig.\,\ref{figTPAevolution}(a) we plot the output profile of the generated CS in  presence (for K=0.04) and in absence (K=0) of TPA. The complete evolution of CS is captured in Fig.\,\ref{figTPAevolution}(b) where an initial oscillation is evident. Note that, in case of a standard Kerr soliton, the peak amplitude [$A(z)$] decays adiabatically due to TPA with an attenuation rate of $(1+8KA(0)^2z/3)^{-1/2}$ \cite{Silberberg}. On the contrary, in the case of CS, the amplitude does not decay continuously but attains a stable value for $K<K_{\rm max}$. It is evident from Fig.~\ref{figTPAevolution}(a) that the peak amplitude of stable CS for $K\neq0$ is reduced compared to the case when $K=0$. Numerically it is found that, for the input parameters $\Delta=3$ and $X=|S|^2=3.5$, there is a maximum value of $K=K_{\rm max}=0.05$. We try to understand this critical phenomenon by steady-state CW bistability analysis, MI analysis, and variational method. In the subsequent sections, we derive all the expressions that are modified due to TPA.

\subsection*{Homogeneous steady-state solutions}
\noindent  
In the presence of TPA ($K \neq 0$),  the steady-state ($\partial u/\partial t =0$), homogeneous  ($\partial u/\partial\tau=0$) solution of Eq.~\eqref{LL} satisfies  the following cubic equation
\begin{equation}
X= (1+K^2)Y^3 - 2(\Delta -K)Y^2 +(\Delta^2 +1)Y, \label{OBtpa} 
\end{equation}
where $Y=|u_s|^2$ and $X=|S|^2$ are the intracavity  and  CW driving power, respectively. It is evident that the steady-state and homogeneous solution of the LLE now depends on the TPA parameter $K$. In absence of TPA ($K=0$), Eq.~\eqref{OBtpa} is converted to  the well-known cubic equation of dispersive optical
bistability for unperturbed LLE, $X= Y^3 - 2\Delta Y^2 +(\Delta^2 +1)Y$ \cite{grelu}.

\subsection*{CW bistability analysis}
\begin{figure*}[t!]
\begin{center}
\epsfig{file=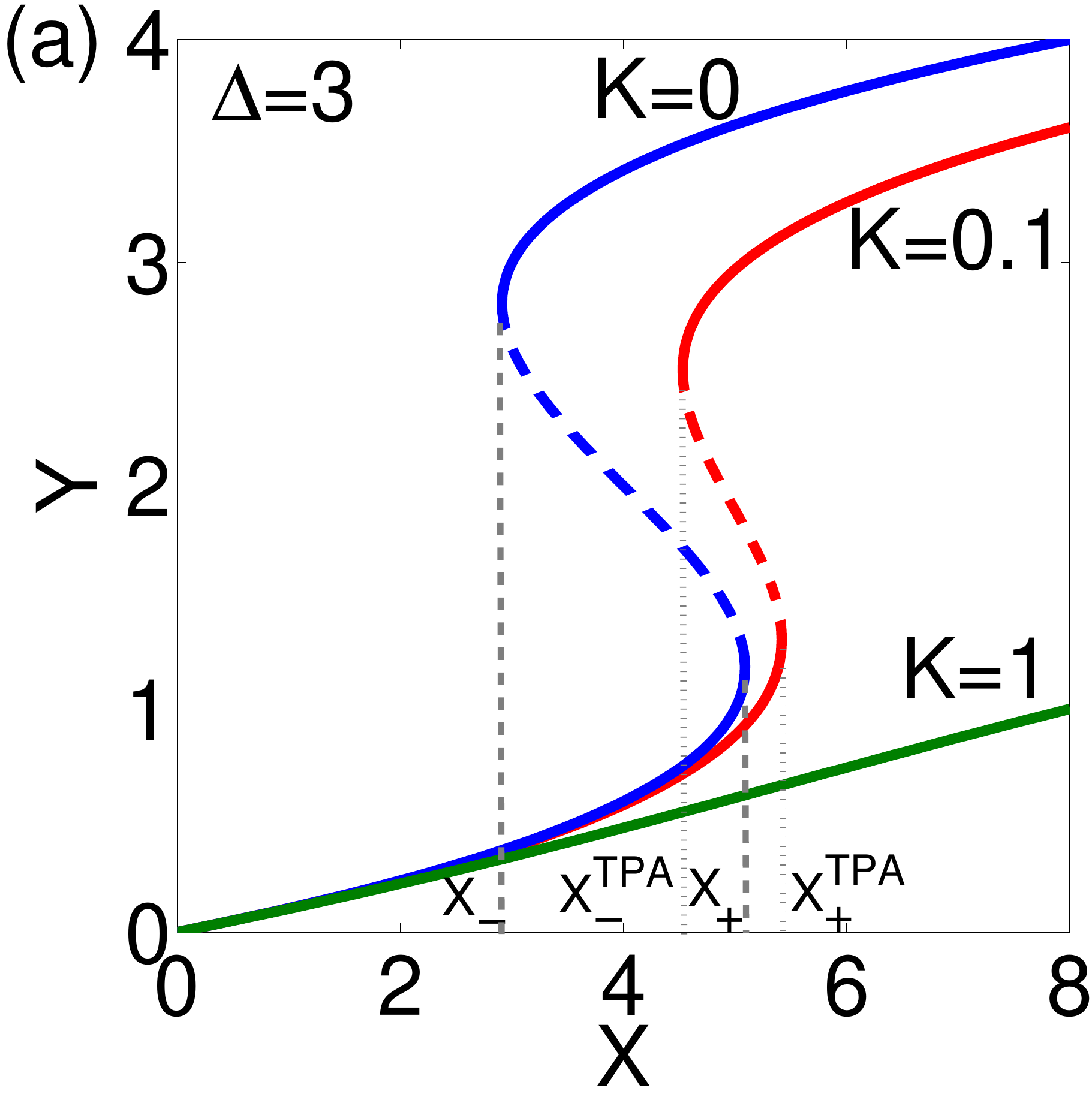,trim=0.0in 0.0in 0.0in 0.0in,clip=true, width=44.0mm}
\epsfig{file=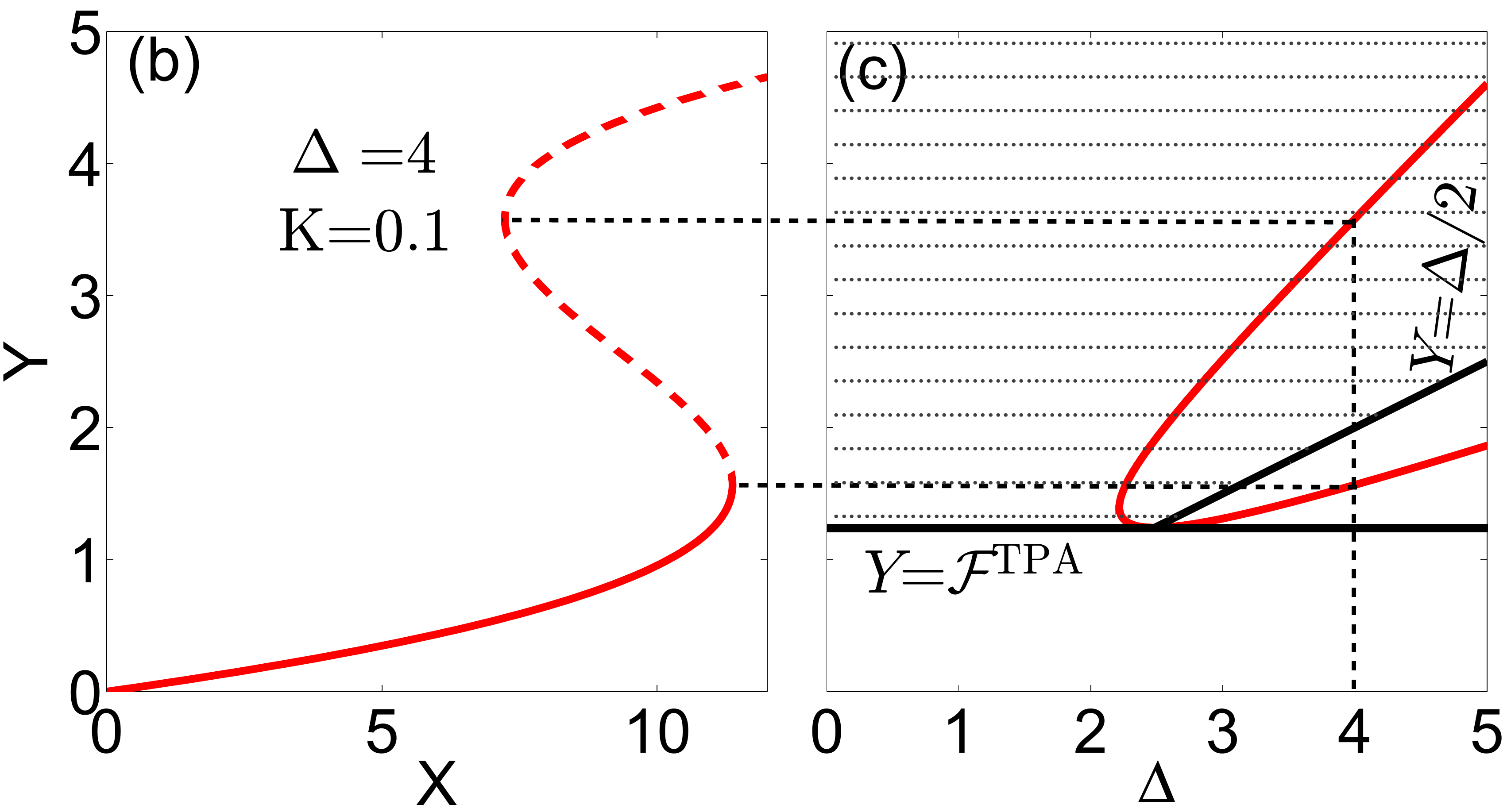,trim=0.0in 0.0in 0.0in 0in,clip=true, width=82mm}
\epsfig{file=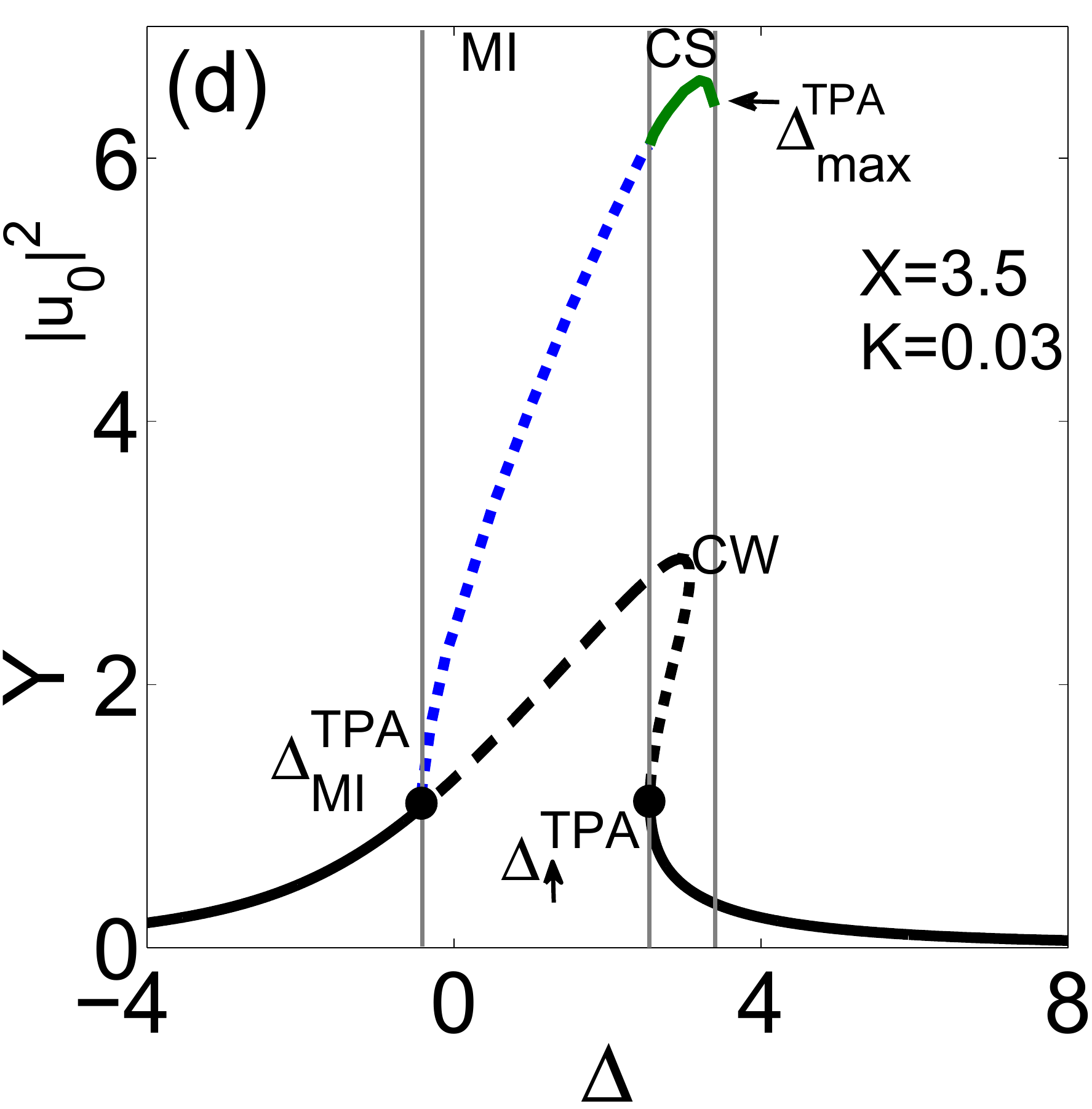,trim=0.0in 0.0in 0.0in 0.0in,clip=true, width=43mm}
\epsfig{file=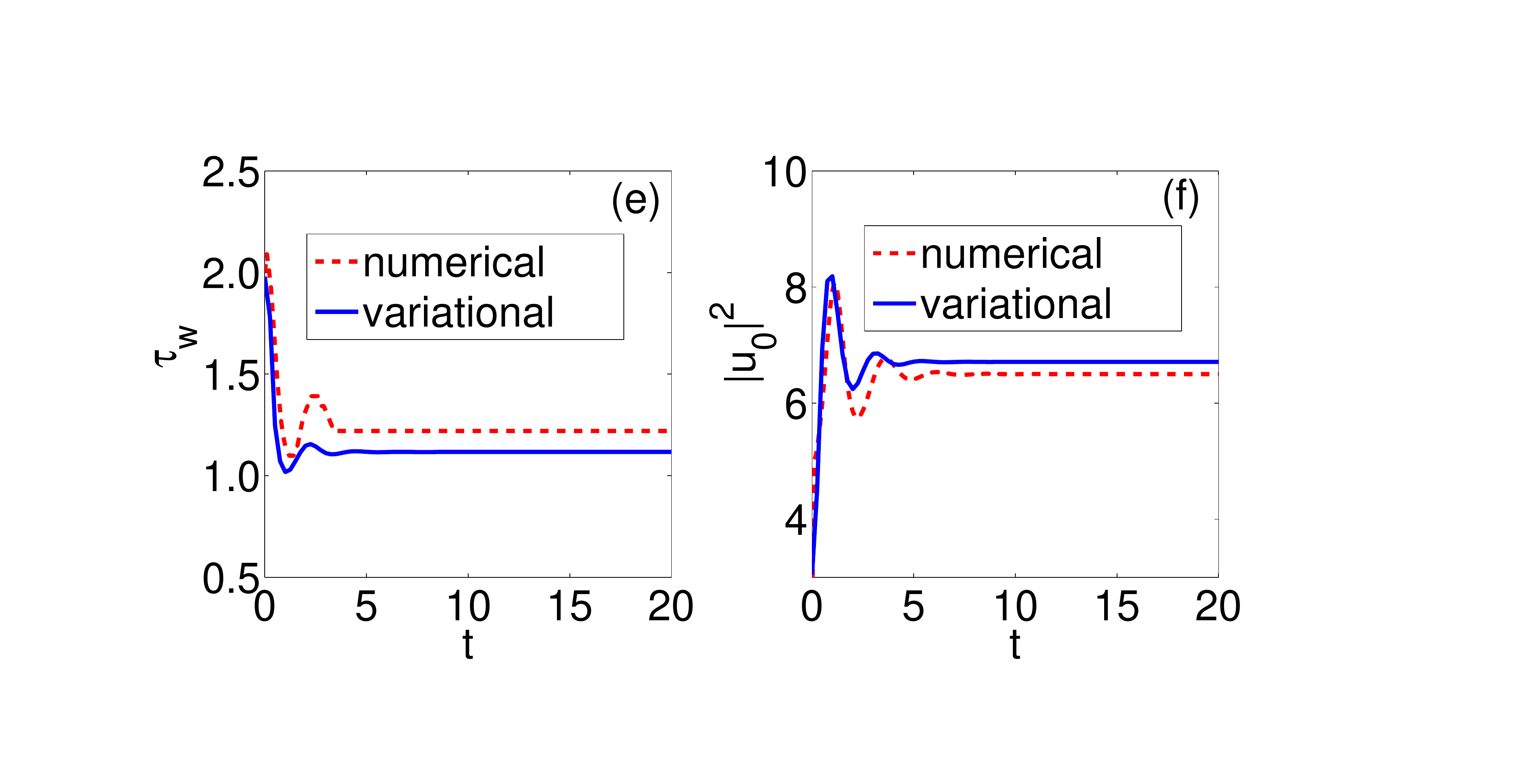,trim=0in 0in 0in 0in,clip=true, width=90mm}
\epsfig{file=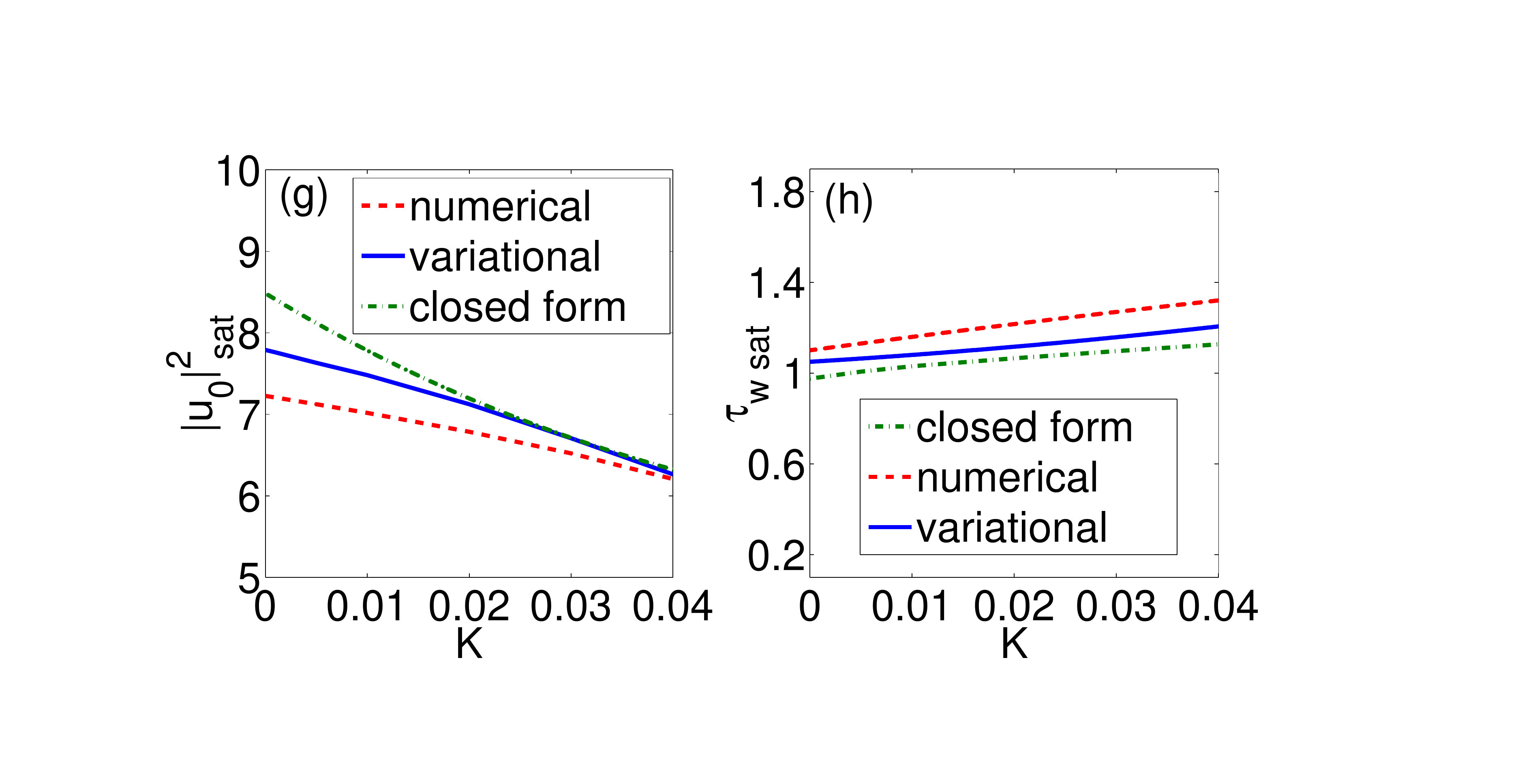,trim=0in 0in 0in 0in,clip=true, width=88mm}
\caption{(Color online) (a) Kerr bistability in $(X,Y)$ parameter space in the case of homogeneous field for three different values of $K$. The turning points are shown by $X_\pm$ (for $K=0$) and $X_\pm^{\rm TPA}$ (for $K\neq0$). (b) Kerr bistability in $(X,Y)$ parameter space, where the dashed parts are unstable. (c) The region of the intracavity MI for anomalous dispersion. The modulationally unstable region is indicated by the shaded area. (d) The steady-state CW intracavity power (black curve), peak intensity of the MI patterns (dashed blue curve), and the peak intensity of CSs (solid green curve) as a function of $\Delta$.
The variation of (e) temporal pulse width $\left(\tau_w=2\eta^{-1}\right)$ and (f) peak intensity $\left(|u_0|^2=E\eta/2\right)$ over the round-trip time $t$ for $K=0.03$. Saturated peak intensity (g) and temporal pulse width (h) as a function of TPA coefficient. Eqs.\,\eqref{TPA_LL_closed_int} and \eqref{TPA_LL_closed_tw} gives the closed form saturated peak intensity and temporal pulse width respectively.}\label{figTPA1}
\end{center}
\end{figure*}

\vspace{-0.0cm} 
\noindent  
Optical bistability analysis in passive Kerr resonators is useful to retrieve different important properties, such as threshold conditions of pump power and the pump detuning that initiates the stable pattern structures \cite{coen}.  In addition to this, the bistability analysis facilitates the understanding of the impact of TPA on the steady-state solution of the LLE. This analysis is important as the coexistence of patterned and CW solutions results in the formation of CS \cite{grelu}. The threshold value of detuning  that initiates the optical bistability can be evaluated from  Eq.\,\eqref{OBtpa}, by setting $dX/dY=0$. For non-vanishing TPA, the steady-state intracavity power $Y_{\pm}$ is given as
\begin{align} \label{Ypm_tpa} 
Y_{\pm}^{\textrm {TPA}}=\frac{2(\Delta-K)\pm \sqrt{(\Delta- K )^2-3(K\Delta+1)^2}}{3(1+K^2)}.
\end{align}
It is easy to show from Eq.\,\eqref{Ypm_tpa} that, the onset of optical bistability for unperturbed LLE $(\Delta_c=\sqrt{3})$  \cite{Haelterman, TH-DM} is modified in presence of TPA as 
\begin{align} \label{Deltapm_tpa} 
\Delta_{\pm}^{\textrm {TPA}}>\left[4K\pm\sqrt{3}(1+K^2)\right]/(1-3K^2).
\end{align}
Unlike the unperturbed case, here we have a range of $K$ values for which the bistability can occur. Equation \eqref{Deltapm_tpa} holds for $\Delta_{+}^{\textrm {TPA}}$ with $K<1/\sqrt{3}$\,, and $\Delta_{-}^{\textrm {TPA}}$ with $K>\sqrt{3}$\,.
The turning points on the bistability curve ($X_{\pm}$) can be calculated as a function of $\Delta$ and have the following form
\begin{align}
\hspace{-0.0cm}
&X_{\pm}^{\rm TPA}=\frac{2}{27(1+K^2)^2}\left[(\Delta-K)\right.\left\{ (\Delta-K)^2 \right. \nonumber\\ &\left. \hspace{0.3cm} + 9(K\Delta+1)^2 \right\} \left. \pm\sqrt{(\Delta- K )^2-3(K\Delta+1)^2}^3  \right].  
\end{align}
In the absence of TPA ($K=0$), this equation reduces to the well-known expressions of up-switching ($X_+$) and down-switching ($X_-$) input powers \cite{grelu}. Also, we can express the Eq.\,\eqref{OBtpa} in terms of detuning $\Delta$ in the presence of TPA as
\begin{align} \label{dddpm_tpa} 
\Delta=Y\pm \sqrt{X/Y -(1+KY)^2}.
\end{align}
In Fig.\,\ref{figTPA1}(a) we plot the steady-state CW response [Eq.\,\eqref{OBtpa}] for three distinct cases $K=0$ (unperturbed), $0.1,$ and $1$. The curve for $K=0.1$ falls within the range of $K<1/\sqrt{3}$ and the bistability is evident with $\Delta>\Delta_{+}^{\rm TPA}(\approx2.22)$. There is no bistability for $K=1$ as it is in the forbidden range $1/\sqrt{3}<K<\sqrt{3}$. On the other hand for $K>\sqrt{3}$, even though the bistability occurs for $\Delta <\sqrt{3}$, the intracavity power and the input power have to be negative to achieve this, which is not physical. In Fig.\,\ref{figTPA1}(a) we also show the up-switching and down-switching points (turning points) for $K=0$ and $K=0.1$. The dotted curves in between  turning points represent the unstable region, where no stable solution can be found. The interpretation of this instability is discussed in detail in the following section using the MI analysis. 

\subsection*{Modulation-instability analysis}
\noindent 
Due to the interplay between dispersion and nonlinear Kerr effect a CW field spontaneously breaks up into a periodic structure through the MI dynamics. Under MI analysis we introduce the ansatz in the form, $u(t,\tau)= u_s +a_+(t)e^{i\Omega\,\tau}+ a_-(t)e^{-i\Omega\,\tau} $ which we insert in Eq. \eqref{LL} that leads to the matrix equation of side-band amplitudes $a_{+}$ and $a_{-}$. The MI gain which is the eigenvalue values of the matrix  $\mathcal{M}$  can be evaluated from eigenvalue equation $|\mathcal{M} - \Lambda I|=0$ (see Appendix B). In the case of TPA, the intracavity MI gain ($\Lambda^{\rm TPA}$) with dimensionless sideband frequency ($\Omega$) is given by
\begin{align} \label{MI_tpa}
\Lambda^{\rm TPA}= -(1+2KY) \pm\sqrt{Y^2(1+K^2)-{\widetilde{\delta}_{\rm TPA}}^2},
\end{align}
where $\widetilde{\delta}_{\rm TPA}=(2Y +\delta_2\Omega^2-\Delta)$.  An instability arises when  $\Lambda^{\rm TPA}$  becomes real positive. In the case of homogeneous perturbations $(\Omega=0)$ this can be achieved when 
\begin{align}\label{MI_tpa15}
\sqrt{4Y\Delta-\Delta^2+(K^2-3)Y^2}\ge 1+2KY.
\end{align}
Note that, the simplification of Eq.\,\eqref{MI_tpa15} is identical to the negative slope of the CW bistability curve (intermediate branch) which corresponds to the unstable region. So, we can conclude from MI analysis that under steady-state CW case the lower and upper branches are always stable, while the intermediate branch (negative slope) is always unstable [shown in Fig.\,\ref{figTPA1}(a)].

\begin{figure}[t!]
\begin{center}
\epsfig{file=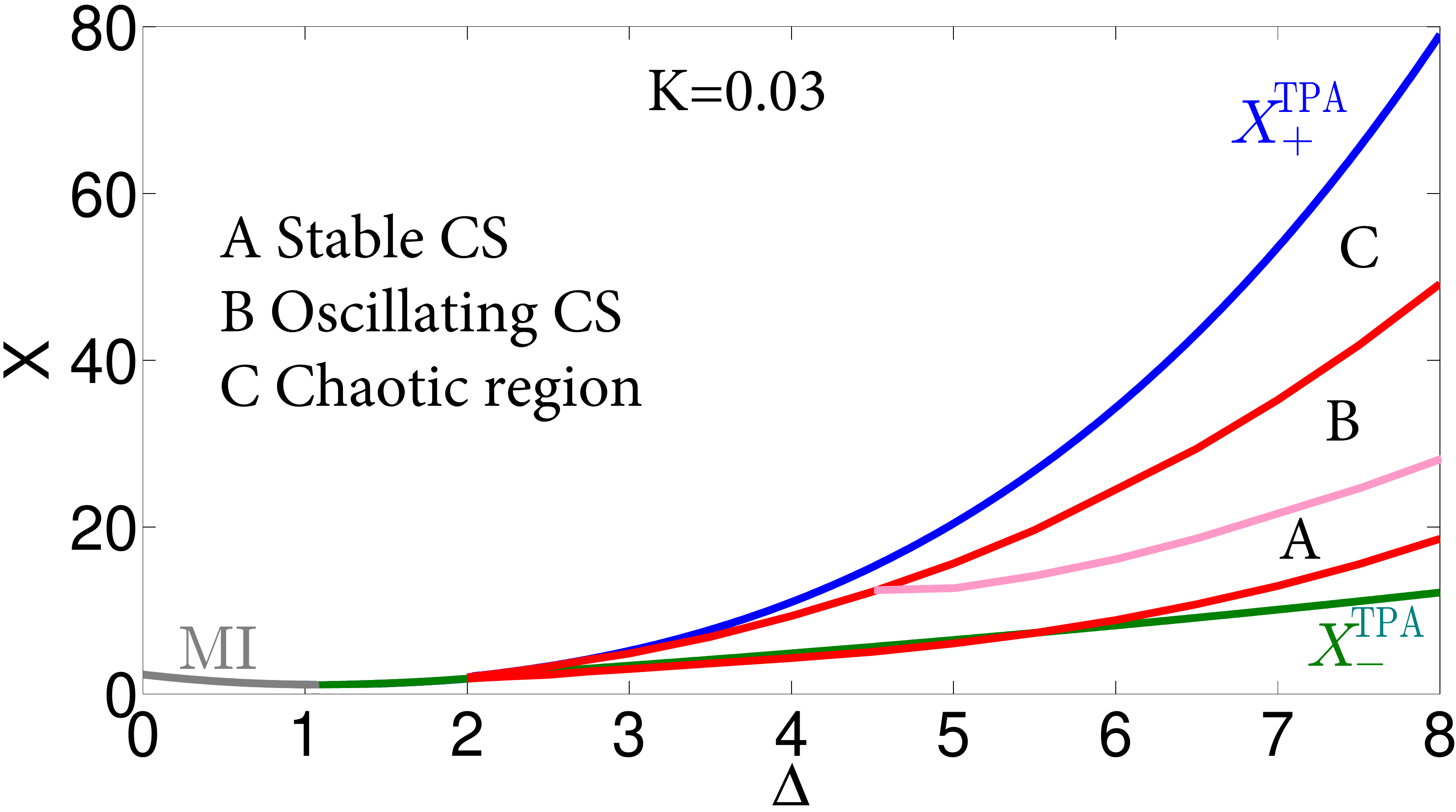,trim=0.0in 0.0in 0.0in 0.0in,clip=true, width=80mm}
\caption{(Color online) Different dynamical regimes of operation and the transition lines in the LLE in the case of TPA.}\label{figTPA_DX}
\end{center}
\end{figure}

In the case of periodic perturbation to the intracavity field ($\Omega\neq0$), with $\delta_2=-1$, the MI gain is obtained when the threshold condition corresponding to $\Lambda^{\rm TPA}=0$ is satisfied \cite{TH-DM}. In the presence of TPA ($K\neq 0$), 
we calculate the threshold conditions that have to be satisfied in order to arise the intracavity MI, which take the form:
\begin{align}
Y\ge \mathcal{F}^{\rm TPA} ~{\rm and} ~Y\ge \Delta/2, ~{\rm with}~\mathcal{F}^{\rm TPA}=\frac{2K+\sqrt{K^2+1}}{1-3K^2}.\nonumber
\end{align}
Note that if we neglect the TPA coefficient ($K=0$), the threshold conditions of MI become $Y\ge 1$ and $Y\ge \Delta/2$, which is the usual condition of MI for unperturbed LLE \cite{grelu}. In Fig.\,\ref{figTPA1}(b) we plot the bistability curve in ($X,\,Y$) parameter space for $\Delta=4$ and $K=0.1$. The modulationally unstable region (shaded region) is shown in Fig.\,\ref{figTPA1}(c) and the correlation is drawn with Fig.\,\ref{figTPA1}(b). Unlike the case of Fig.\,\ref{figTPA1}(a), here periodic perturbations ($\Omega\neq 0$) make the upper branch unstable as shown by the dashed portion of the bistability curve in Fig.\,\ref{figTPA1}(b). In Fig.\,\ref{figTPA1}(d) we plot the bistability curve (black curve) $\Delta$ vs $Y$ for $X=3.5$ with a TPA coefficient of $K=0.03$. The peak intensity of MI patterns as a function of $\Delta$ is shown by the dashed blue line. The minimum value of detuning where MI starts is calculated from Eq.\,\eqref{dddpm_tpa} with the threshold MI condition as $\Delta_{\rm MI}^{\rm TPA}=\mathcal{F}^{\rm TPA}-\sqrt{X/\mathcal{F}^{\rm TPA} -(1+K\mathcal{F}^{\rm TPA})^2}$. We also plot the peak intensity of stable CS as a function of $\Delta$ (green curve) and indicate the limiting points of the CS branch $\Delta_{\uparrow}^{\rm TPA}$ \cite{coen} and $\Delta_{\rm max}^{\rm TPA}$.  The theoretical expression of $\Delta_{\rm max}^{\rm TPA}$ is calculated analytically exploiting the variational analysis in the following section. In Fig.\,\ref{figTPA_DX} we show an attractor chart of the LLE \cite{Leo13,parrarivas14} in the parameter space ($X,\,\Delta$). In this plot we illustrate different dynamical regimes that are separated by transition lines as indicated in the figure.

%
%

\subsection*{Perturbative analysis}
\noindent We adopt a semianalytical variational technique to study the complex dynamics of a perturbed CS under TPA. The treatment allows us to explore the problem with greater insights. In this process, we derive a set of coupled ODEs that predicts the change in amplitude and pulse width of CS under TPA. In Fig.\,\ref{figTPAevolution}(a) we plot the shape of CS obtained numerically by solving the LLE for $K=0$ (blue dotted line) and $K=0.04$ (gray dotted line). The corresponding variational ansatzes are also depicted in the same plot. It is observed that the peak amplitude of the input pulse initially experiences an oscillation that is damped down to a steady value to form a CS. This feature is illustrated in Figs.\,\ref{figTPAevolution}(b),  \ref{figTPA1}(e), and \ref{figTPA1}(f). The variational results (solid lines) agree well with the numerical simulations (dashed lines). In the absence of TPA, the saturated peak intensity of CS is given as $|u_0|^2_{sat}\approx \pi^2S^2/4$ [see Eq.\,\eqref{A11}]. It is obvious that this saturated value will reduce in presence of TPA ($K\neq0$).  In Figs.\,\ref{figTPA1}(e) and \ref{figTPA1}(f) we plot the evolutions of temporal width and peak intensity of CS for nonvanishing $K$. The red dotted lines indicate the results obtained from the full numerical solution of LLE, whereas the blue solid lines represent the variational outcome that we achieve by solving the set of coupled ODEs [Eq.~\eqref{var6}-\eqref{var10}] for $K\neq0$. During both calculations, we have kept other perturbations zero. The coupled equations become more useful if we decouple them with proper approximations. We can obtain the saturated values of the intensity and temporal width under TPA ($K<K_{\rm max}$) by approximating Eqs.\,\eqref{var6} and \eqref{var10}, which takes the following form:
\begin{align} 
|u_0|_{sat}^2& \approx \frac{1}{4}\left[\mathcal{A}^{1/3}/K - 2 \mathcal{A}^{-1/3} \right],\label{TPA_LL_closed_int}\\
\tau_{w\,sat}& = 2\sqrt{2}/|u_0|_{sat}, \label{TPA_LL_closed_tw}
\end{align}
with $\mathcal{A}=3\pi S K^2 +\sqrt{8K^3+9\pi^2S^2K^4}$. In Figs.\,\ref{figTPA1}(g) and \ref{figTPA1}(h) we demonstrate how the saturated intensity ($|u_0|^2_{sat}$) and pulse width ($\tau_{w \,sat}$) depend on the TPA parameter $K\,(<K_{\rm max})$. Note that, the variational results [solid line in plot (g)] disagree more for small values of $K$. The accuracy of variational treatment largely depends on the proper choice of initial ansatz function. Actually, the results become more accurate when the ansatz fits well with the actual pulse shape. If we carefully study Fig.\,\ref{figTPAevolution}(a), we find that the proposed ansatz (solid line) agrees better with the actual pulse shape (dashed line) for larger $K$ value. Hence it is expected that we get better agreement for a relatively larger $K$ which is indeed the case.

The variational analysis [Eqs.\,\eqref{var6},\,\eqref{var9} and \eqref{var10}] can also determine the theoretical limit of $\Delta_{\rm max} (=\pi^2 X/8)$ upto which the unperturbed CS can sustain [See Eq.\,\eqref{A13}]. In the presence of TPA, using the same analysis we can derive an equation 
\begin{align} \label{tpa_maxparam}
\Delta\left(\frac{4}{3}K\Delta +1 \right)^2= \Delta_{\rm{max}},
\end{align}
from where we can calculate the maximum values of $\Delta$ and $K$. For a fixed K, the maximum detuning $\Delta_{\rm max}^{\rm TPA}$ is calculated by putting $\Delta=\Delta_{\rm max}^{\rm TPA}$ in Eq.\,\eqref{tpa_maxparam} and solving the cubic equation, that provides a new theoretical limit of the detuning upto which the CS can sustain under TPA
\begin{align} 
\Delta_{\rm max}^{\rm TPA}=\frac{(K-\mathcal{B}^{1/3})^2}{4K^2 \mathcal{B}^{1/3}},
\end{align}
where $\mathcal{B}=K^3+18\Delta_{\rm max}K^4+6 \sqrt{\Delta_{\rm max}K^7(1+9\Delta_{\rm max}K)}$.
For a given $K (=0.03)$ and  an external power $X(=3.5)$, the maximum detuning comes out to be, $\Delta_{\rm max}^{\rm TPA}=3.36$. For the same set of ($K,\,X$) we run our simulation and confirm that CS exists for the maximum value of detuning $\Delta \approx 3.4$ which agrees closely with our theoretical prediction. Similarly, for fixed values of $\Delta$ and $X$ we can calculate the maximum value of the TPA coefficient ($K_{\rm max}$) upto which CS can sustain. By solving the quadratic equation for $K_{\rm max}$ [Eq.\,\eqref{tpa_maxparam}] we obtain
\begin{align}
K_{\rm max}=-\frac{3}{4\Delta^2}\left( \Delta -\sqrt{\Delta_{\rm max}\Delta} \right).
\end{align} 
For a given $X=3.5$ and $\Delta=3$ the theoretical value of $K_{\rm max}=0.0499$ upto which CS can exist, that matches with our numerical results of $K=0.05$.

\section{Impact of free carriers on cavity soliton}
\noindent The wide transparency and large material nonlinearity of Si make it an advantageous photonic component for integrated optical devices such as a microcavity. However, in a realistic Si-based microresonator, when pumped below 2.2 $\mu m$ wavelength, TPA becomes dominant. It leads to the generation of free carriers in the form of electron-hole pairs that introduces additional loss (FCA) and also change the refractive index through FCD. We take into account these effects in the LLE where the rate equation of FC is coupled to it [Eq.\,\eqref{LL}]. The stability and dynamics of CS is expected to be affected by the FC perturbation. The full numerical simulation reveals that the spatially accumulated FC density ($\phi_c$) over multiple round trips modifies the stability condition of CS excitation. In order to grasp the role of free carriers on the stability of CS, we numerically solve the coupled LLE for $\Delta = 3$, $X = |S|^2 = 3.5$, and $\mu=3.7741$  which is the realistic value calculated in \cite{Lau}. The accumulated FC density over the first round trip is calculated as $\phi_c=0.18$. The CS is found to be formed for this value of FC density. But the stable structure of CS is disrupted in the second round trip when FC density reaches to $\phi_c=0.36$. The accumulation of the FC density over successive round trips is calculated through the boundary condition $\phi_c(t,-\mathcal{T_R}/2)=\phi_c(t+\Delta t,+\mathcal{T_R}/2) $ \cite{Lau}. In the following sections, using steady-state CW bistability analysis we derive threshold values of the system parameters that are necessary to excite CS.

\subsection*{Homogeneous steady-state solutions}
\begin{figure*}[t!]
\begin{center}
\epsfig{file=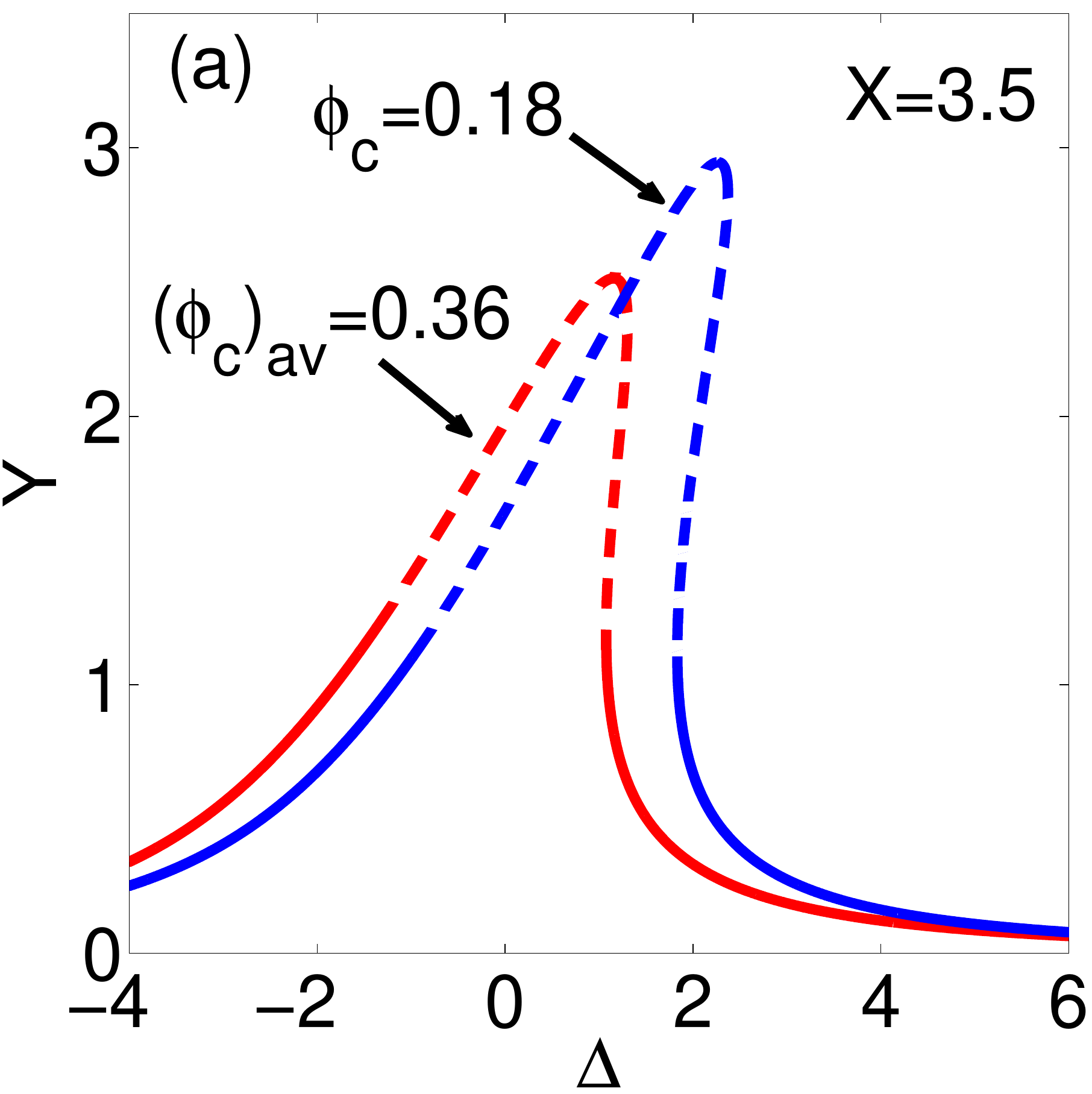,trim=0.0in 0.0in 0.0in 0.0in,clip=true, width=42.5mm}
\epsfig{file=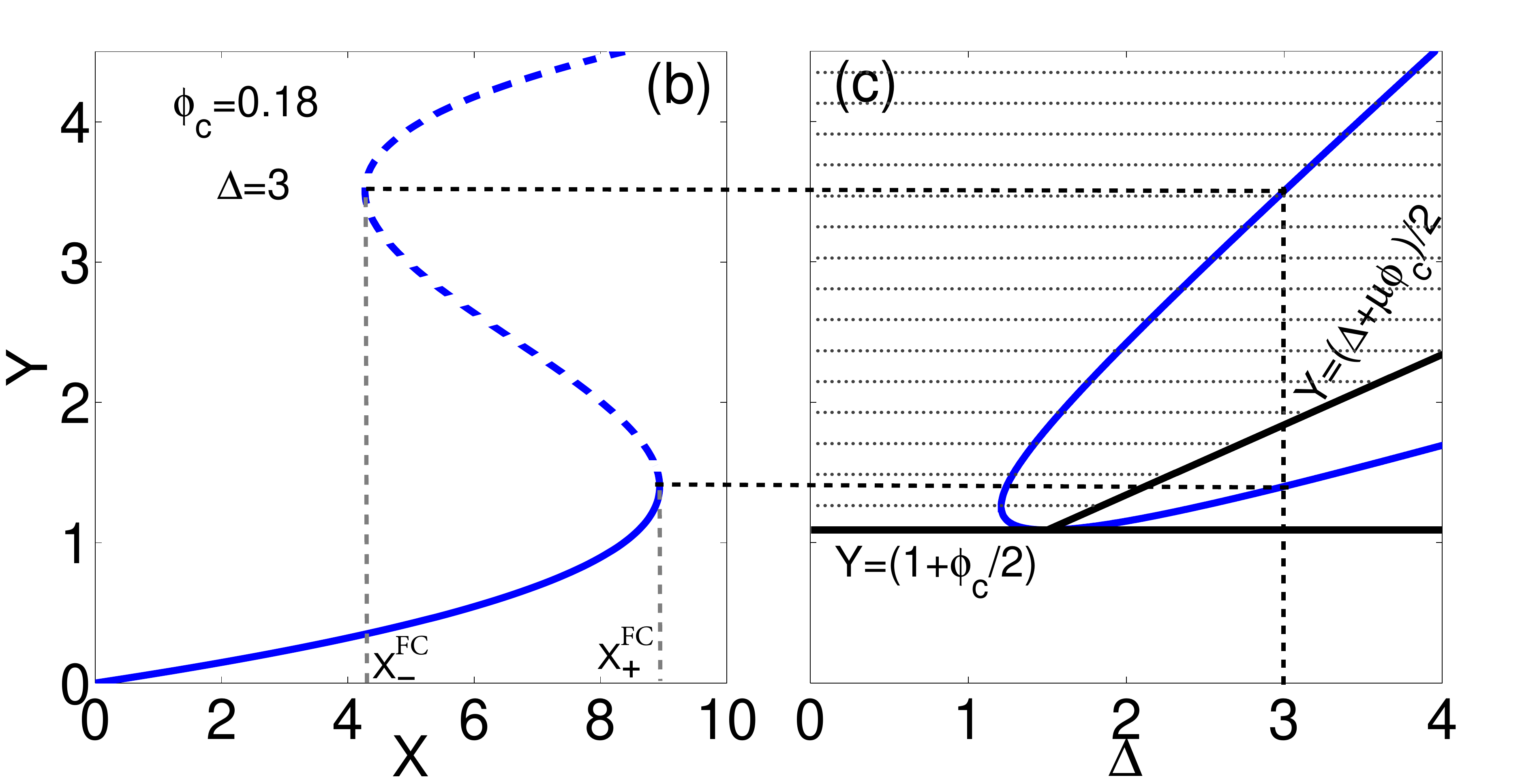,trim=0.0in 0.0in 0.0in 0.0in,clip=true, width=83.5mm}
\epsfig{file=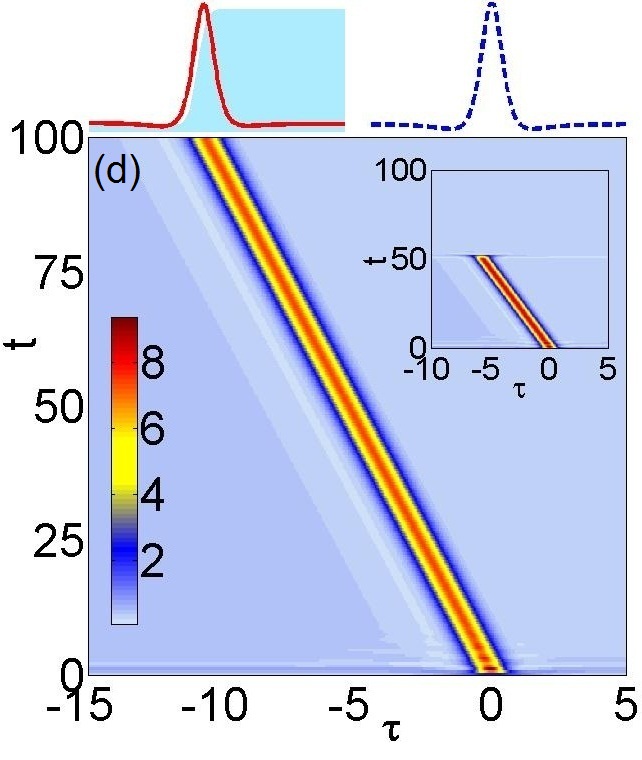,trim=0in 0.00in 0in 0.0in,clip=true, width=43mm}
\vspace{0em}
\epsfig{file=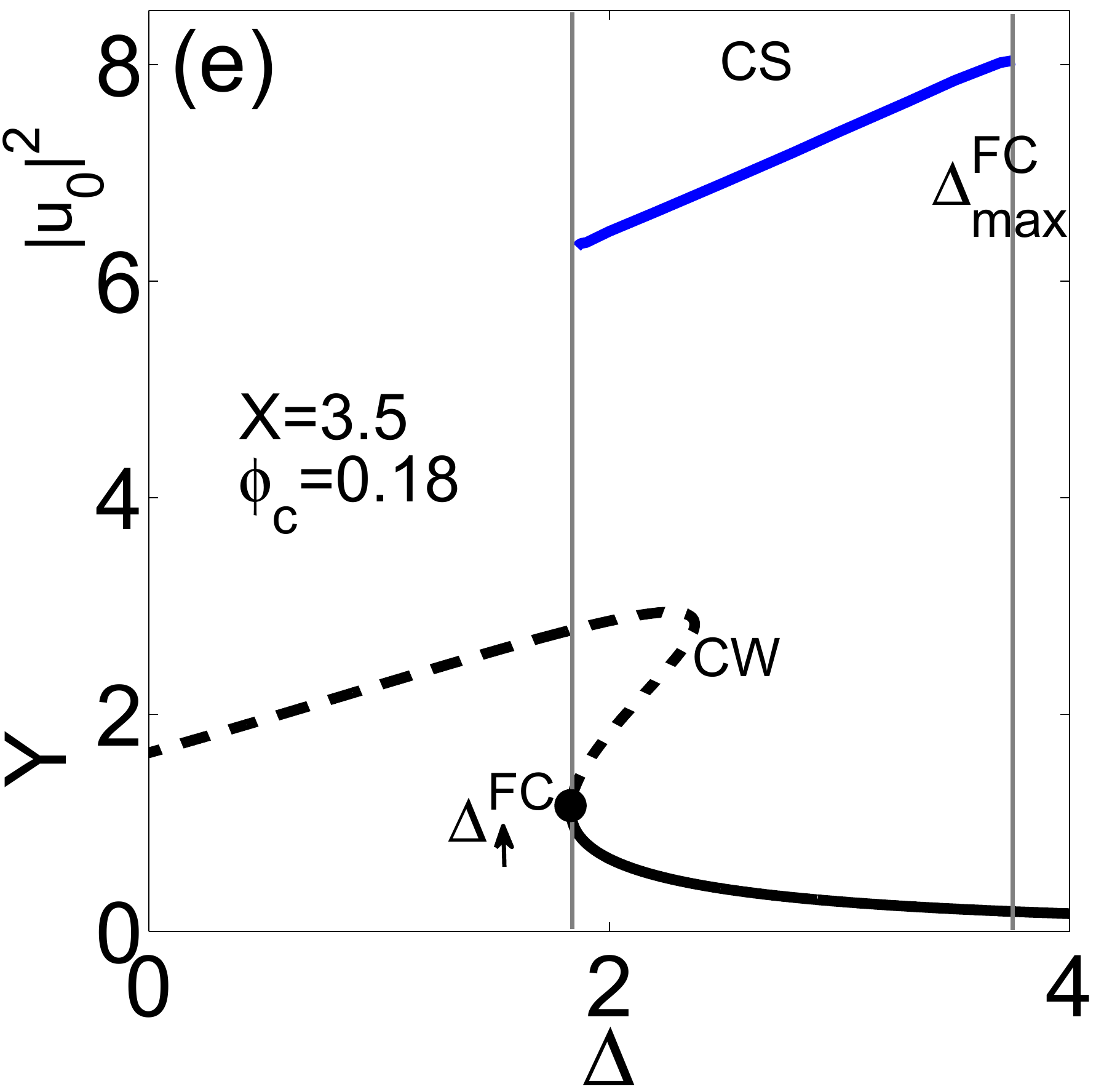,trim=0.0in 0.0in 0.0in 0.0in,clip=true, width=42.0mm}
\epsfig{file=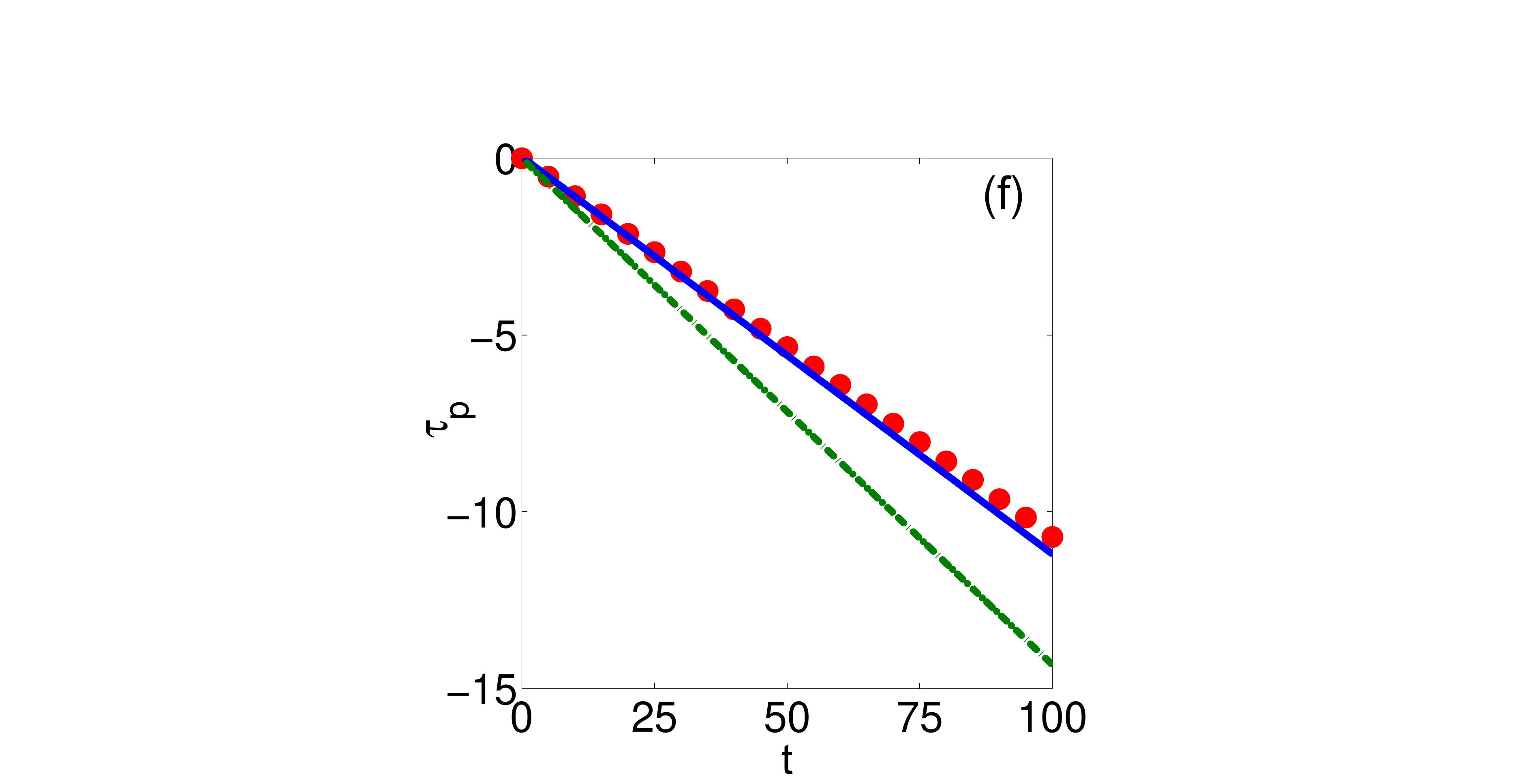,trim=0in 0.0in 0in 0in,clip=true, width=44.5mm}
\epsfig{file=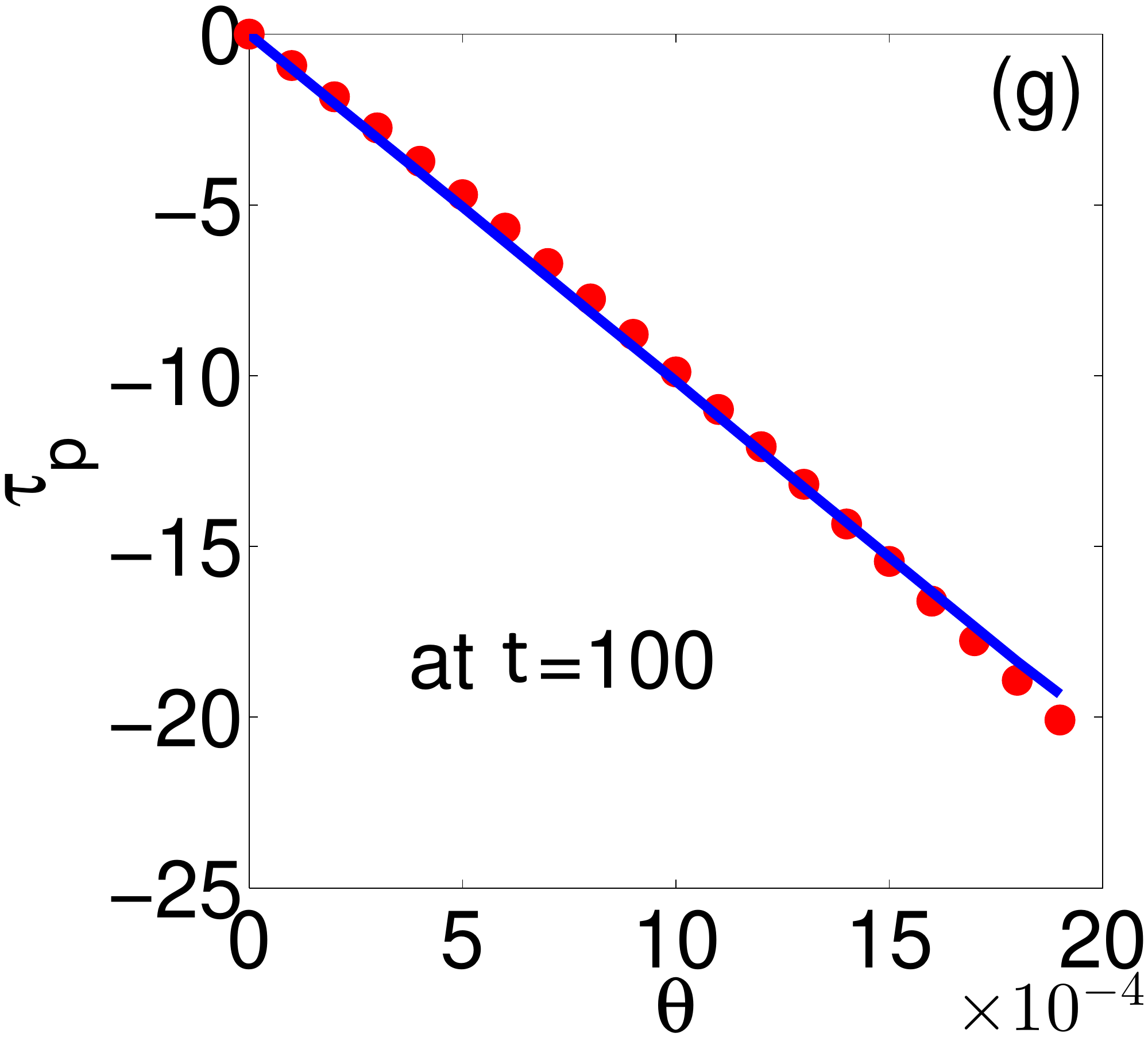,trim=0in 0.0in 0in 0in,clip=true, width=47mm}
\epsfig{file=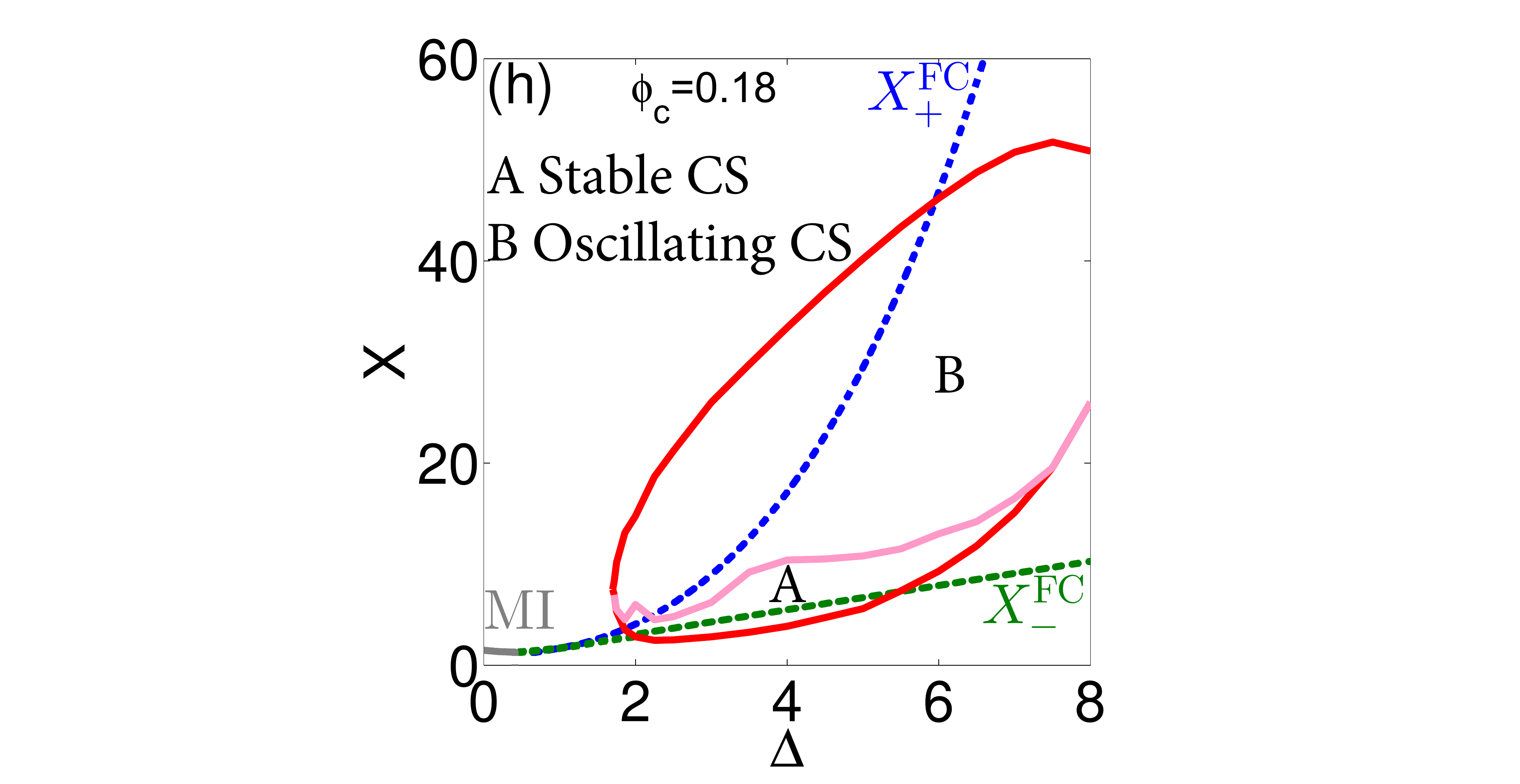,trim=0.0in 0.0in 0.0in 0.0in,clip=true, width=42.5mm}
\vspace{0em}
\caption{(Color online) Kerr bistability in (a) $(\Delta,Y)$ parameter space for two values of $\phi_c$, and in (b) $(X,Y)$ parameter space. (c) The region of the intracavity MI for anomalous dispersion. The modulationally unstable region is indicated by the shaded area.
(d) Temporal evolution of CS profile for realistic $\theta=0.0011$ and $\mu=3.7741$ with the parameters $\Delta=3$ and $X=3.5$ for single round trip. The inset gives the same evolution for multiple round trips. The unperturbed (dotted line) and perturbed (solid line) temporal CSs are also shown on the overhead, also the normalized $\phi_c$ is shown at $t=100$ for $\tau_c=10^{-3}$ by the light-green-shaded area on the top. 
(e) The CW response (black curve) and the peak intensity of CSs (solid green curve) as a function of $\Delta$. (f) The variation of temporal delay of CS as a function of t is shown for a single round trip. The red circles give numerical data, whereas the solid blue line gives variational prediction. The closed-form of the temporal delay [Eq.~\eqref{closed_FC_t}] is also plotted by a green dot-dashed line. (g) Comparison between the  numerical (red circles) and variational (solid blue line) results of the temporal delay of CS as a function of $\theta$. (h) Different dynamical regimes of operation in the LLE for $\mu=7.7741$ and $\theta=0.0011$.}\label{figFC}
\end{center}
\end{figure*}

\noindent 
The circulating electric field inside the Si-microresonator accumulates free carriers. Under FC generation ($\phi_c \neq 0$) the intracavity power $Y=|u_s|^2$ relates the driving field $X$ as [steady-state CW solution of Eq.\,\eqref{LL}]
\begin{align} \label{OBfc}
&X=Y^3 - 2(\Delta+\mu\phi_c) Y^2 \nonumber \\ &\hspace{2.3cm}+ \left\{ (\Delta+\mu\phi_c)^2  + \left(1+{\phi_c}/{2}\right)^2 \right\}Y.
\end{align}
This steady-state CW solution is a cubic equation of $Y$ which generally a function of $\Delta$ and $X$, and in case of perturbation due to free carriers it depends on $\mu$ and $\phi_c$. In Figs.\,\ref{figFC}(a) and \ref{figFC}(b) we plot Eq.\,\eqref{OBfc}  for $Y$ as a function of $\Delta$ (for fixed $X$), and as a function of $X$ (for fixed $\Delta$). From these figures, it is evident that depending on the values of system parameters the intracavity power becomes multivalued and we get three solutions in total. Using CW bistability analysis and MI analysis we calculate the expressions of critical parameters, turning points, and also the stable and unstable branches of bistability curves.

\subsection*{CW bistability analysis}
\noindent  
Similar to the TPA perturbation, here the impact of free carriers on the steady-state CW solutions of the LLE is analyzed which is essential for the existence of the CS.
Exploiting the cubic equation [Eq.\,\eqref{OBfc}] we can calculate the analytical  expression of the threshold detuning $\Delta_c^{\rm FC}$ which initiates the optical bistability. $\Delta_c^{\rm FC}$ can be calculated from the expression of the steady-state intracavity power $Y_{\pm}$ (by setting $dX/dY=0$)
\begin{align} \label{Ypm_fc} 
Y_{\pm}^{\textrm {FC}}=\frac{2(\Delta+\mu\phi_c)\pm \sqrt{(\Delta+\mu\phi_c)^2-3(1+\phi_c/2)^2}}{3}.
\end{align}
Here we find that the threshold detuning  depends on the FCD coefficient ($\mu$) and density of the generated free carriers ($\phi_c$) as
\begin{align} \label{Deltapm_fc} 
\Delta_{\pm}^{\textrm {FC}}>\pm\sqrt{3}\left(1+\phi_c/2 \right) -\mu\phi_c \,.
\end{align}
For a fixed FCD coefficient $\mu\approx3.774$, bistability occurs for $\Delta_{+}^{\textrm {FC}}$ with $\phi_c<\phi_{c\, \rm max}\left[= \sqrt{3}/(\mu-\sqrt{3/2})\right]$. The expression of turning points can be calculated in the presence of free carriers as 
\begin{align}
&X_{\pm}^{\rm FC}=\frac{2}{27}\left[(\Delta+\mu\phi_c)\{ (\Delta+\mu\phi_c)^2 +9 (1+\phi_c/2)^2  \}   \right. \nonumber\\ &\left. \hspace{2cm}\pm\sqrt{(\Delta+\mu\phi_c)^2-3(1+\phi_c/2)^2}^3  \right].  
\end{align}
It is also possible to represent Eq.\,\eqref{OBfc} in terms of $\Delta$, which takes the following form 
\begin{align} \label{dddpm_fc} 
\Delta=(Y-\mu\phi_c)\pm \sqrt{X/Y -(1+\phi_c/2)^2}.
\end{align}

\subsection*{Modulation-instability analysis}
\noindent 
Under FC generation, the intracavity MI gain ($\Lambda^{\rm FC}$) is given by the eigenvalues of the matrix $\mathcal{M}$ as (see Appendix B)
\begin{align} \label{MI_fc}
\Lambda^{\rm FC}=& -\left(1+\phi_c/2 \right) \pm\sqrt{Y^2- \widetilde{\delta}_{\rm FC}^2},
\end{align}
where $\widetilde{\delta}_{\rm FC}=\left(2Y-\Delta-\mu\phi_c+\delta_2\Omega^2 \right)$. Depending on the system parameters, instability arises for real positive values of $\Lambda^{\rm FC}$ \cite{TH-DM}. For homogeneous  perturbations $(\Omega=0)$, the instability arises when 
\begin{align}
\sqrt{Y^2-\left(2Y-\Delta-\mu\phi_c \right)^2}\ge 1+\phi_c/2.
\end{align}
This expression provides the unstable solutions of the intermediate branch of the CW bistability curve in the presence of free carriers. Now, for the periodic perturbation to the intracavity field ($\Omega\neq0$), with $\delta_2=-1$, the MI arises (by setting $\Lambda^{\rm FC}=0$) when $Y$ satisfies the conditions $Y\ge (1+\phi_c/2) ~{\rm and} ~Y\ge (\Delta+\mu\phi_c)/2$.
The minimum value of detuning where MI starts is also calculated from Eq.\,\eqref{dddpm_fc} as $\Delta_{\rm MI}^{\rm FC}=(1+\phi_c/2-\mu\phi_c)-\sqrt{X/(1+\phi_c/2) -(1+\phi_c/2)^2}$.

In Figs.\,\ref{figFC}(a) and \ref{figFC}(b), the unstable intermediate and upper branches are illustrated by the dashed portion of the bistability curves. From these plots, it is evident that the accumulation of free carriers ($\phi_c$) due to multiple round trips shifts the bistability curve and modifies the threshold value of the parameters. For $\phi_c=0.18$, which is less than $\phi_{c\,\rm max}\,(\approx0.596)$, the bistability can occur for relatively small detuning $\Delta>\Delta_+^{\rm FC}(=1.2086)$. The turning points are also depicted in Fig.\,\ref{figFC}(b) as $X_{\pm}^{\rm FC}$. In Fig.\,\ref{figFC}(c) we show the modulationally unstable region (shaded region), and draw a correlation with Fig.\,\ref{figFC}(b). The accumulation of free carriers ($\phi_c$) due to multiple round trips shifts the bistability curve that limits the generation of CS. We illustrate this phenomenon in Fig.\,\ref{figFC}(d) by plotting the evolution of the CS which we obtain by solving Eq.\eqref{LL} for $\phi_c \neq0$. In the inset, we demonstrate how stable CS ceases to exist after a few round trips with larger FC density. In Fig.\,\ref{figFC}(e) we plot the bistability curve (black curve) $\Delta$ vs $Y$ for $X=3.5$ with $\phi_c=0.18$, where the dashed portion gives the unstable region. In the same plot, we depict the variation of the peak intensity of stable CS as a function of $\Delta$ (solid blue curve). We also indicate the onset detuning $\Delta_{\uparrow}^{\rm TPA}=1.86$, and  the maximum detuning $\Delta_{\rm max}^{\rm TPA}=3.8$ upto which CS can exist in the presence of free carriers.

\subsection*{Perturbative analysis}
\noindent 
The influence of free carriers on the dynamics of CS can be understood if we solve the variational equations Eq.\,\eqref{var6}-\eqref{var10} containing only the FC term ($\theta \neq 0$). FCA reduces the steady amplitude of the CS and the pulse energy $E$ saturates to $E_{sat}\approx\sqrt{2}\,\pi S/\left(1+ \pi \mu \theta S/2\sqrt{2}\right)$ [see Eq.\,\eqref{Afc_energy}]. The CS accelerates due to the FC induced index change and experiences frequency blueshifting. The frequency blueshifting due to the FCD is approximately calculated as $\Omega_{p_{\,\rm {FC}}}^{sat}\approx\sqrt{2}\,\mu\theta (E_{sat}\,\eta_{sat})^{5/2}/15\pi S$ [see Eq.\,\eqref{Afc_freq}]. Note that, it is difficult to obtain a clear spectral shift $\Omega_p$ of CS in the presence of FC through numerical simulation as the side-wing of the spectrum is destroyed. However, the temporal acceleration of CS, as a consequence of spectral blueshift, is efficiently calculated by exploiting the variational results. The expression of the temporal shift is approximated as [see Eq.\,\eqref{Afc_taup}]
\begin{equation}\label{closed_FC_t}
\tau_p(t) \approx\left(-2\,\Omega_{p_{\,\rm {FC}}}^{sat}-\frac{7}{72}\theta E_{sat}^2\right)t.
\end{equation} 
The group delay of CS ($\tau_p$) over slow time $t$ is plotted in Fig.\,\ref{figFC}(f), where the numerical data (solid dot) is in good agreement with the variational result (solid-blue line) that we obtain by solving the set of ODEs Eq.~\eqref{var6}-\eqref{var10}. The analytical expression of temporal shift [Eq.~\eqref{closed_FC_t}] is also depicted in the same plot through dotted-green line. Note, the closed form we derive in Eq.~\eqref{closed_FC_t} is based on certain approximation (see Appendix A) and this approximation might lead to the slight deviation. In Fig.\,\ref{figFC}(g) we plot the $\tau_p$ (the group delay of CS) as a function of $\theta$ (FC generation term) at $t=100$, where the variational predictions (solid blue line) match well with the numerical data (solid dots).  Exploiting the results obtained from the variational analysis [Eqs.\,\eqref{var6},\,\eqref{var9} and \eqref{var10}] we try to determine the theoretical limit of $\Delta_{\rm max}^{\rm FC}$, that takes the following form 
\begin{align} \label{FC_maxparam}
\Delta_{\rm max}^{\rm FC}\approx\Delta_{\rm{max}}-&\frac{\theta\eta_{sat}E_{sat}^2}{24}\left(5\mu  \right.\nonumber \\ &\hspace{0.5cm}\left. + \eta_{sat}E_{sat} +\theta\eta_{sat}^2E_{sat}^3/24 \right).
\end{align}
From our approximated closed-form expression [Eq. \eqref{FC_maxparam}] we get $\Delta_{\rm max}^{\rm FC} \approx 4.0$, which is  close to the value that we obtained by full numerical simulation. In Fig.\,\ref{figFC}(h) we illustrate an attractor chart of the LLE \cite{Leo13,parrarivas14} in the parameter space ($X,\,\Delta$) for nonvanishing $\phi_c$. In this plot, we also depict the main bifurcation lines $X_\pm^{\rm FC}$ and different dynamical regimes that are separated by transition lines. It is evident from this figure that the impact of FCA on the CS is significant and the stable CS is generated within a very limited region (region A).

\section{Impact of Intrapulse Raman Scattering on cavity soliton}
\begin{figure*}[tb]
\begin{center}
\epsfig{file=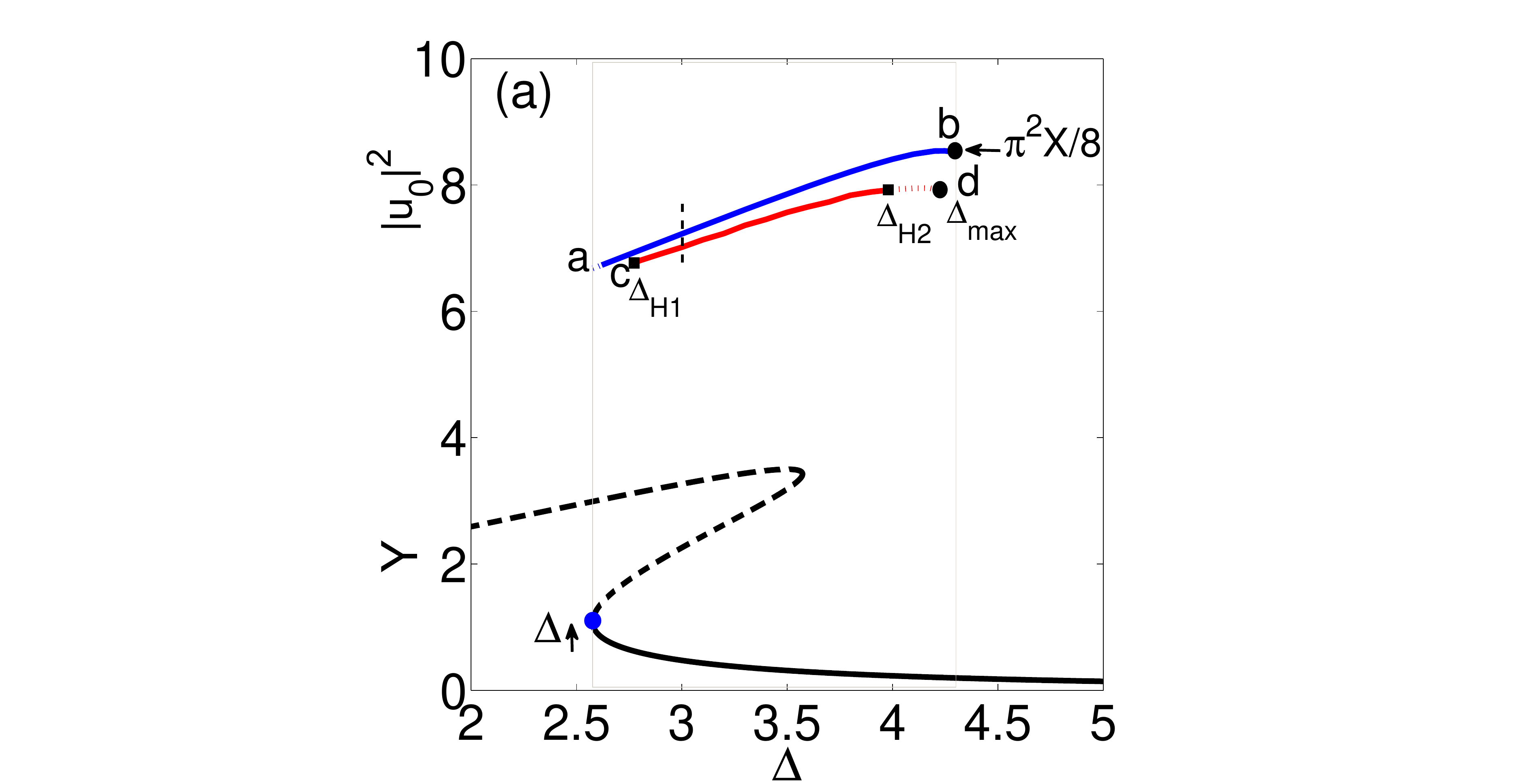,trim=3.35in 0.05in 4.0in 0.4in,clip=true, width=50mm}
\epsfig{file=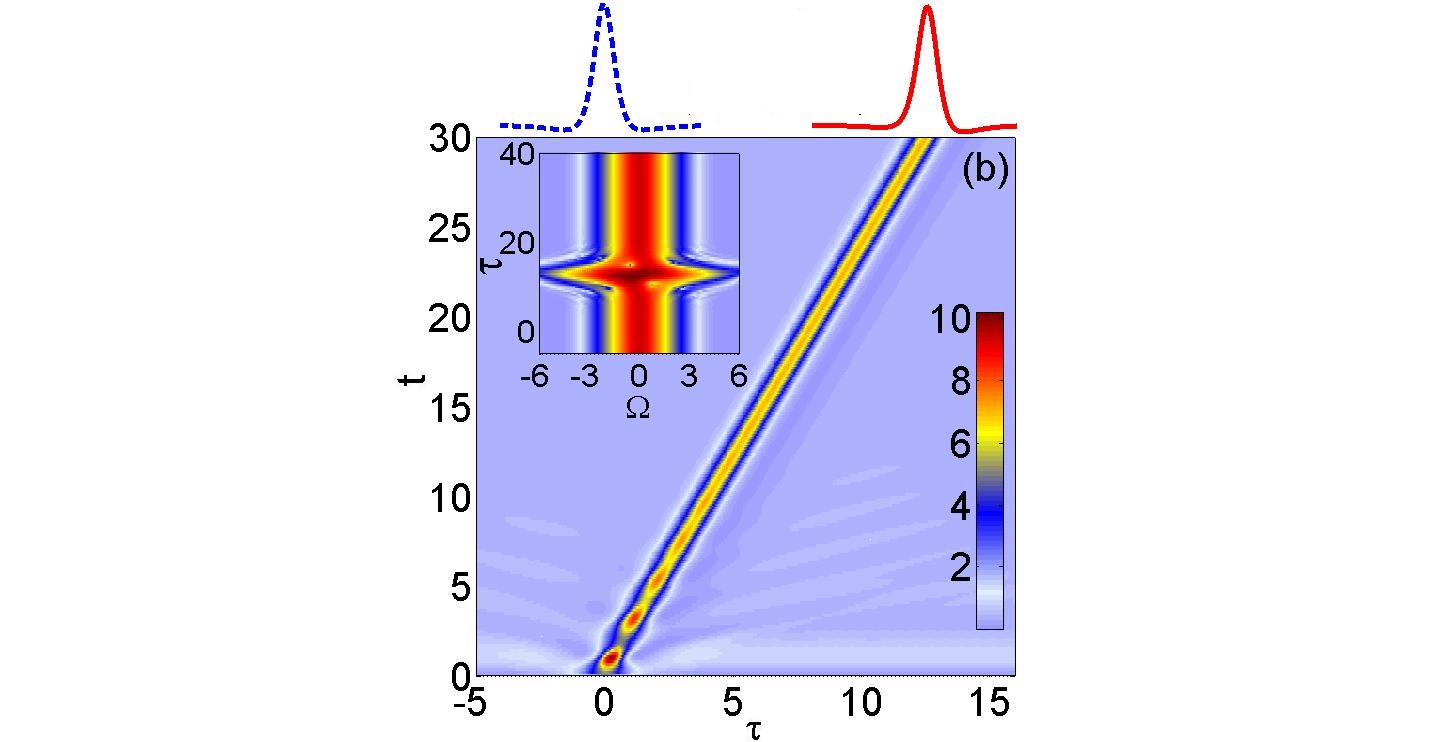,trim=4.0in 0.00in 4.1in 0.0in,clip=true, width=51mm}
\epsfig{file=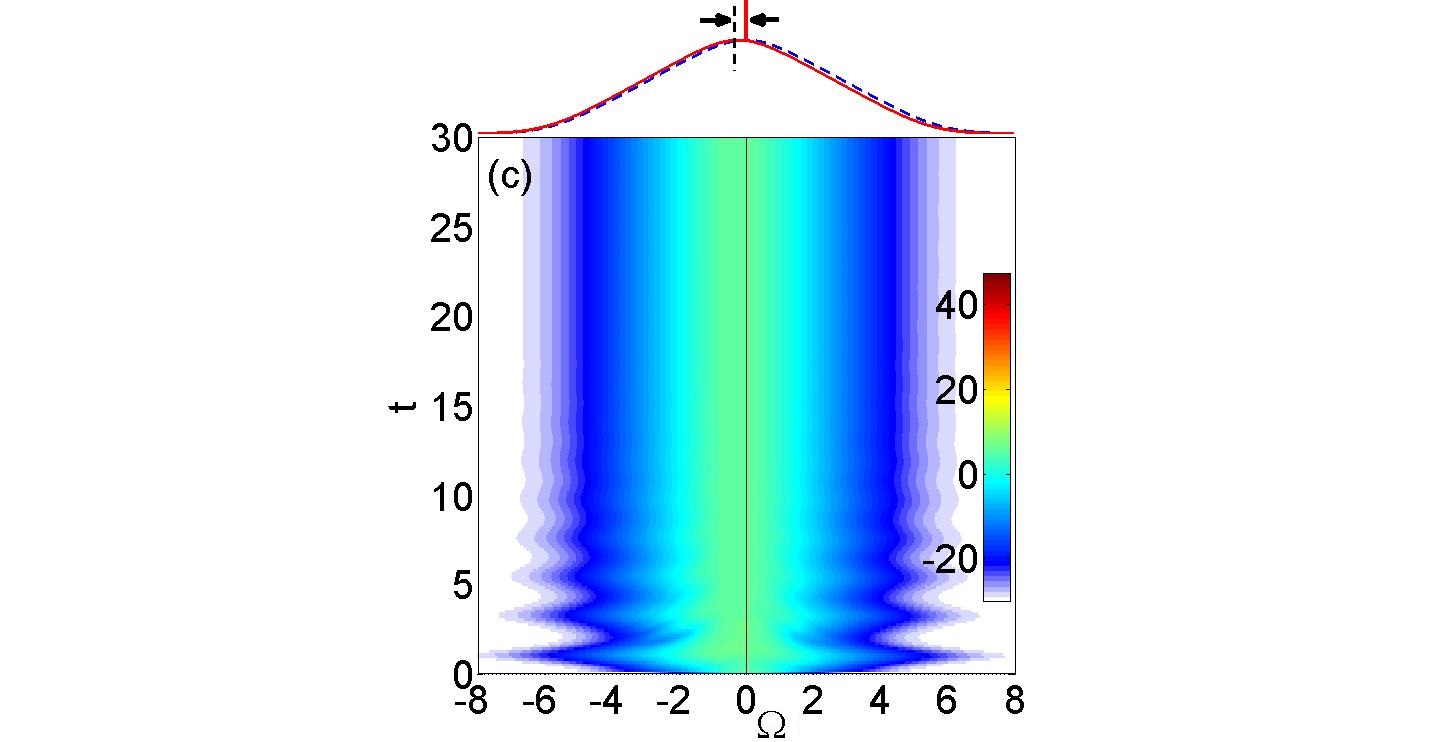,trim=4.0in 0.00in 4.1in 0.0in,clip=true, width=51mm}
\vspace{0.0em}
\epsfig{file=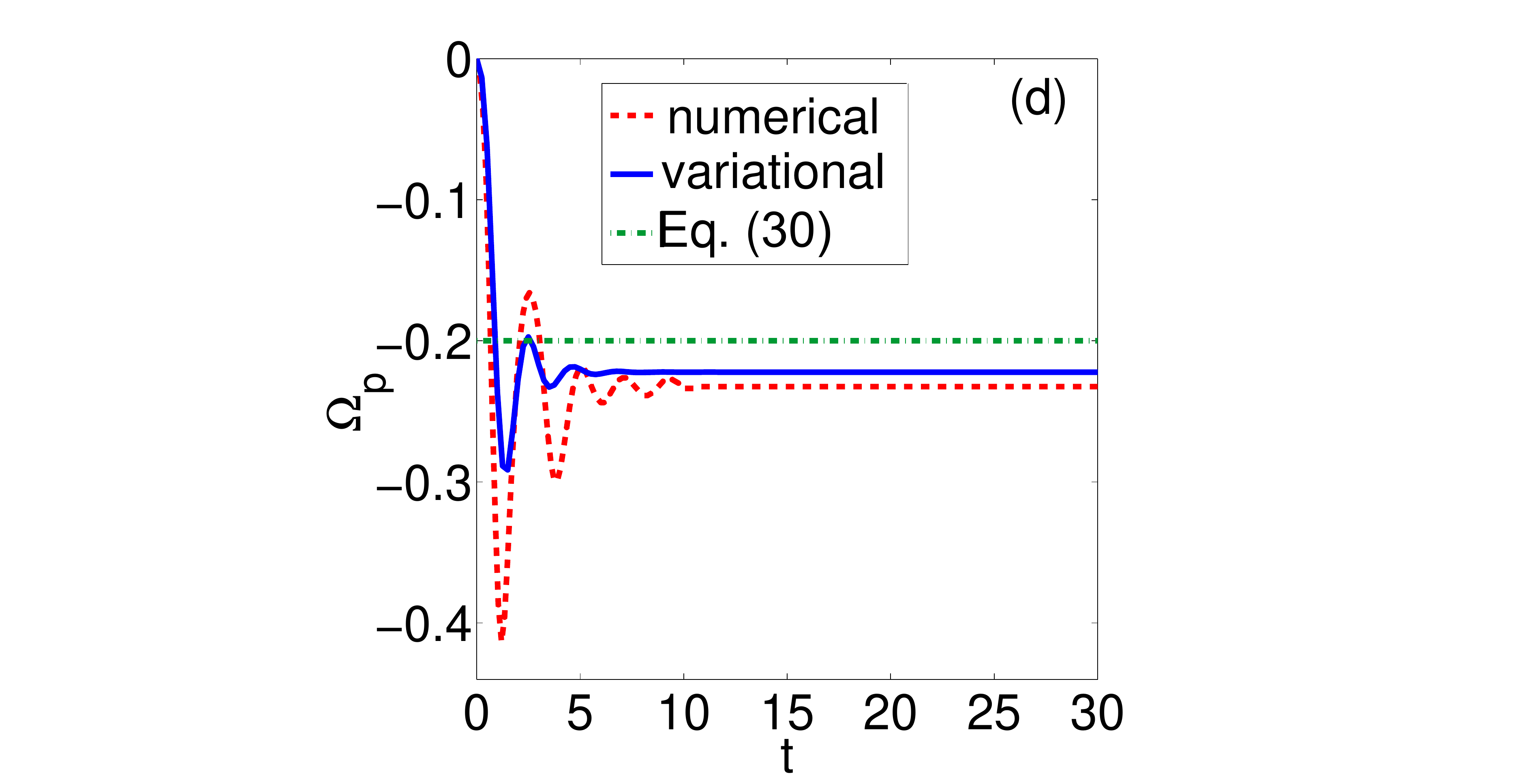,trim=0in 0.00in 0in 0.0in,clip=true, width=51mm}
\epsfig{file=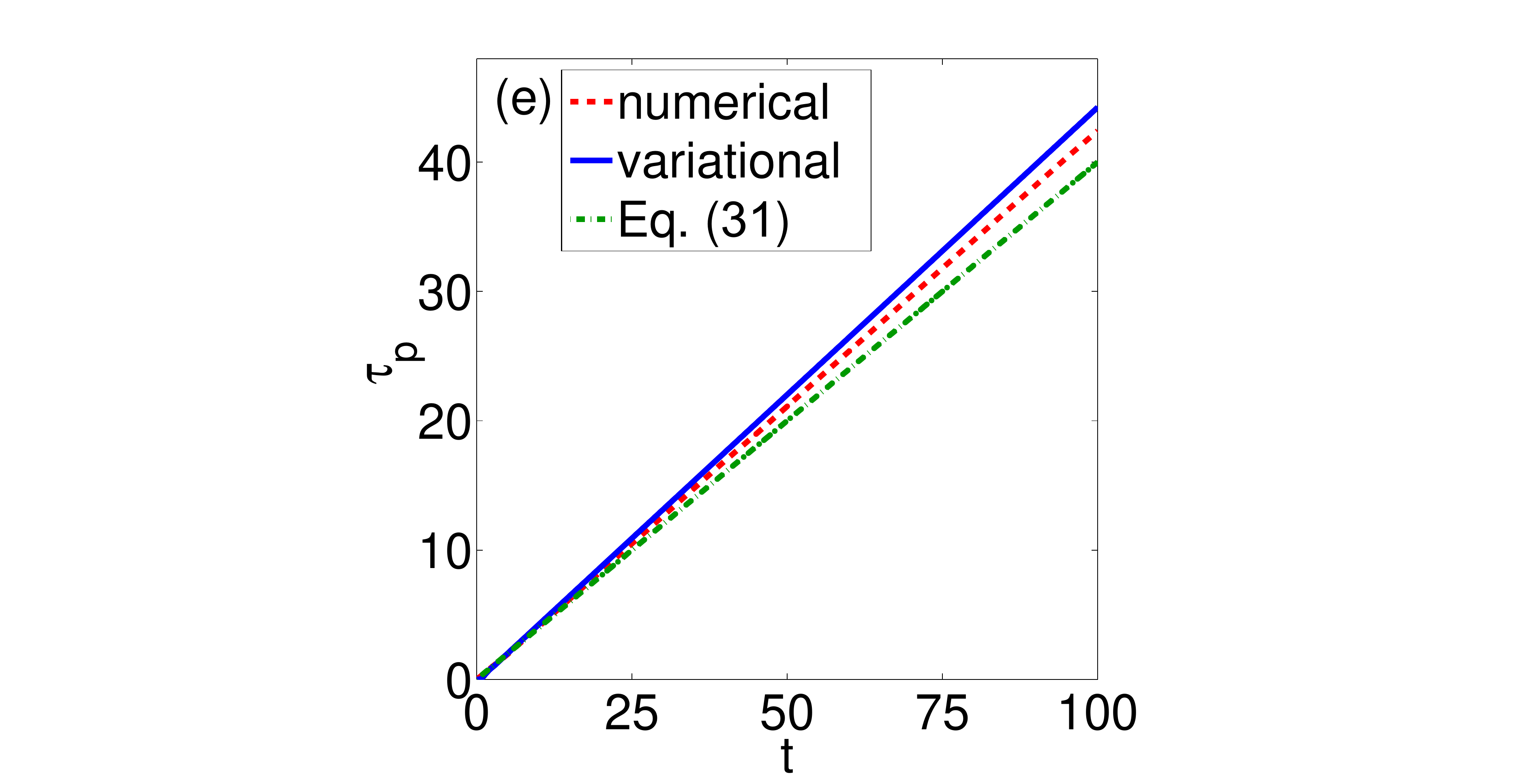,trim=0in 0.00in 0in 0.0in,clip=true, width=49mm}
\epsfig{file=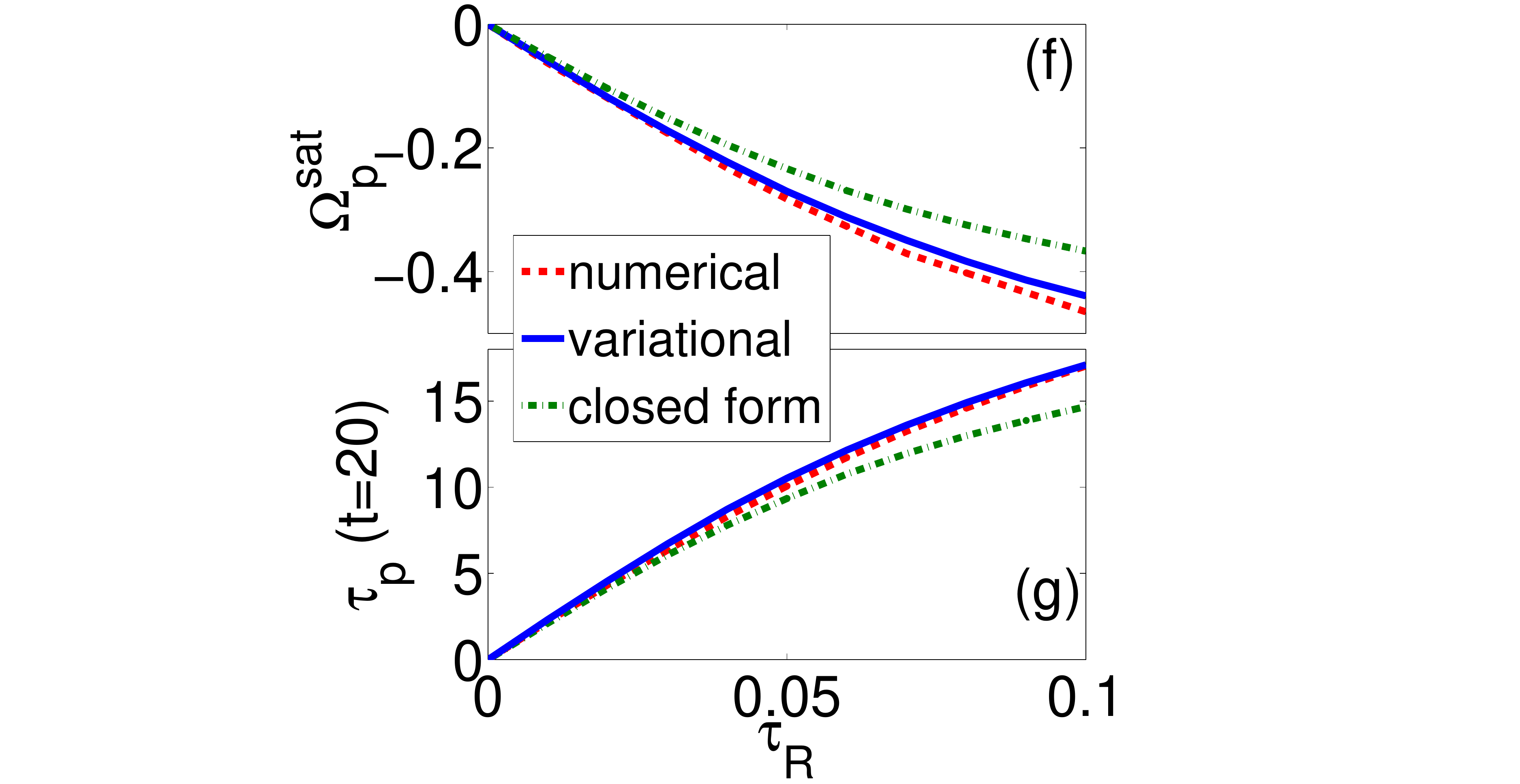,trim=0in 0.0in 0.0in 0.0in,clip=true, width=50.5mm}
\vspace{0.0em}
\caption{(Color online) (a) Peak intensity of the intracavity field as a function of cavity detuning $\Delta$ for $X=3.5$. Black curve represents CW solution, while red c-d and blue a-b curves show CS solutions with and without IRS ($\tau_R=0.04$), respectively. The dashed curves correspond to unstable solutions. The vertical dashed line represents the point where $\Delta=3$. (b) Temporal ($\tau$) and (c) spectral [$\Omega=(\omega-\omega_0)\tau_s$] evolution of CS profiles for $\tau_R=0.04$ with parameters $\Delta=3$ and $X=3.5$. The unperturbed (dotted trace) and  perturbed (solid line) CSs in temporal and spectral domains are also shown on the top of each panel. The inset of (b) gives the spectrogram plot at $t=30$. The variation of (d) frequency shift and (e) temporal delay of CS over the round-trip time t. (f) The saturated frequency  and (g) corresponding temporal delay at $t=20$ as a function of $\tau_R$.}\label{fig-raman}
\end{center}
\end{figure*}

\subsection*{Homogeneous steady-state solutions, stability and existence of CSs} 
\noindent 
The stability of the steady-state solution of the unperturbed LLE is governed by the parameters $\Delta$ and $S$ \cite{coen}. IRS introduces an additional parameter $\tau_R$ which influences the stability and limits the duration and bandwidth of temporal CS \cite{Erkintalo}. The CW bistability condition does not change due to the IRS and gives the identical Eq.~\eqref{OBtpa} with $K=0$. Note that, we perform the MI analysis in the presence of IRS and notice the MI gain (positive real value) does not contain any term related to $\tau_R$. Hence MI calculation for IRS does not provide any new information regarding the stability of the steady-state solution of LLE. In Fig.\,\ref{fig-raman}(a) we plot the CW bistability curve in ($Y,\Delta$) parameter space which has the MI unstable region (dashed portion). To illustrate how IRS affects the stability and existence of CSs, in Fig.\,\ref{fig-raman}(a) we plot $|u_0|^2$ as a function of $\Delta$ at fixed $X=|S|^2=3.5$. In order to get $|u_0|^2$, we numerically solve Eq.\,\eqref{LL} (for IRS perturbation alone) in presence and absence of  $\tau_R$ and obtain two lines as depicted by a-b ($\tau_R=0$) and c-d ($\tau_R=0.04$). CSs exist for both the cases $\tau_R=0$ and $\tau_R\neq0$, when $\Delta>\Delta_{\uparrow}$, the up-switching point. The unstable CS under IRS is stabilized through an inverse Hopf bifurcation at $\Delta_{H1}$ and sustained upto $\Delta_{H2}$ and again becomes unstable \cite{Erkintalo}. With increasing $\Delta$ the output intensity of perturbed CS reduces which compliments the variational results. The CS solutions under IRS cease to exist at a detuning $\Delta_{\rm max}$ which is less than the theoretical limit of unperturbed detuning of $\pi^2X/8$.  The temporal and spectral evolutions of a perturbed CS are shown in Fig.\,\ref{fig-raman}(b) and (c), respectively, for $\Delta=3$ and $X=3.5$. The spectrogram is also shown in the inset of Fig.\,\ref{fig-raman}(b). IRS induces a temporal deceleration and spectral redshift to the CS. The frequency-shift eventually saturates to a steady value \cite{Vahala,Kippenberg} over the round-trip time.

\subsection*{Perturbative analysis}
\noindent In this section we solve the coupled differential equations obtained by the variational approach [Eqs.\,\eqref{var6}-\eqref{var10}] considering IRS as the only perturbation ($\tau_R \neq 0$) and compare the analytical predictions with the results obtained from full numerical simulations of Eq.\,\eqref{LL}. In Figs.\,\ref{fig-raman}(d) and \ref{fig-raman}(e) we plot the evolutions of the frequency shift and temporal position of a CS perturbed under IRS. As shown in Fig.\,\ref{fig-raman}(d) the frequency shift is stabilized followed by an initial oscillation. Interestingly the variational treatment (solid blue line) also captures this initial oscillations with good agreement (red dotted line). The linear temporal shift $\tau_p$ is also well predicted by the variation analysis as demonstrated in Fig.\,\ref{fig-raman}(e). The analytical and numerical studies reveal that the parameters $|u_0|^2$, $\eta$ and $E$ are  stabilized to some fixed values $|u_0|_{sat}^2$, $\eta_{sat}$ and $E_{sat}=2|u_0|_{sat}^2/\eta_{sat}$ within few round trips. Considering the steady frequency down-shifting ${d\Omega_p^{sat}}/{dt}=0$ and assuming ${\rm sech}{\left({\pi \Omega_p^{sat}}/{2 \eta_{sat}}  \right)}\,\cos{\phi}\approx1$ in Eq.\,\eqref{var8}, we can write the expression of the frequency that saturates as
\begin{equation} \label{closed_IRS}
  \Omega_p^{sat}\approx -\frac{2\sqrt{2}}{15}\frac{\tau_R}{\pi S}E_{sat}^{\frac{3}{2}} \,\eta_{sat}^{\frac{7}{2}}.
\end{equation}
The saturated frequency down-shifting leads to a monotonous temporal shift of CS
\begin{equation} \label{closed_IRS_t}
    \tau_p(t) \approx -2\,\Omega_p^{sat}\,t.
\end{equation}
This equation shows that $\tau_p$ varies linearly with slow-time $t$, which is evident in Fig.\,\ref{fig-raman}(e). In the plot the red-dotted line represents the group delay ($\tau_p$) of CS as a function of $t$, which we obtain by solving Eq.\,\eqref{LL} numerically. The variational analysis which is the solution of Eq.\,\eqref{var6}-\eqref{var10} results in a closed match (solid blue line) with numerical data. Further, we try to evaluate $\tau_p$ exploiting Eq.\,\eqref{closed_IRS_t} and depict the result (green dotted-line) in the same plot. In Figs.\,\ref{fig-raman}(f) and \ref{fig-raman}(g) we plot $\Omega_p^{sat}$ and $\tau_p$ at $t=20$ as a function of $\tau_R$. The full variational and approximated analytical results agree with the numerical data.

\section{Impact of lossy phase-modulated driving field on cavity soliton}
\noindent 
Controlled excitation of CS that persists in driven passive cavity systems, is found to be interesting in different applications \cite{JKJ-SC}. The CS can be controlled by a phase-modulated driving field. The selective writing and erasing of CS can be possible for such an arrangement. Considering a complex phase of the driving field, the intracavity field amplitude $u(t,\tau)$ is modeled \cite{LL,coen} as
\begin{equation} 
\frac{\partial u}{\partial t }=  \left[-1 +i\left(|u|^2 - \Delta\right) - i ~sgn(\beta_2) \frac{\partial^2 }{\partial \tau^2 } \right]u +S(t), \label{LL_pd}
\end{equation}
where $S(t)=S_0 \, \exp[(-\rho+i\sigma)t]$, with $\rho$ and $\sigma$ are the pump depletion coefficient and the phase of the pump, respectively. Eq.\,\eqref{LL_pd} is solved numerically for $sgn(\beta_2)=-1$ and the solutions are plotted in Fig.~\ref{figPD}(a) for $\rho=0.005$.  It is obvious that the loss due to the pump depletion affects the existence of CSs.

\subsection*{Homogeneous steady-state solutions}
\begin{figure}[tb]
\begin{center}
\epsfig{file=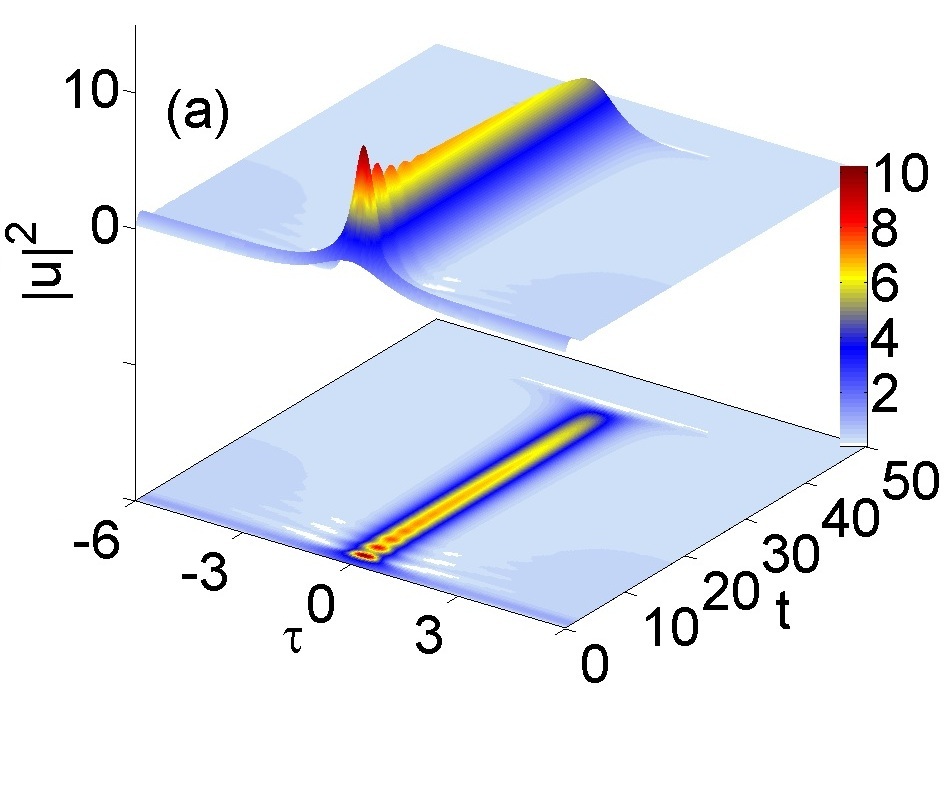,trim=0.0in 0.0in 0.0in 0.0in,clip=true, width=43.4mm}
\epsfig{file=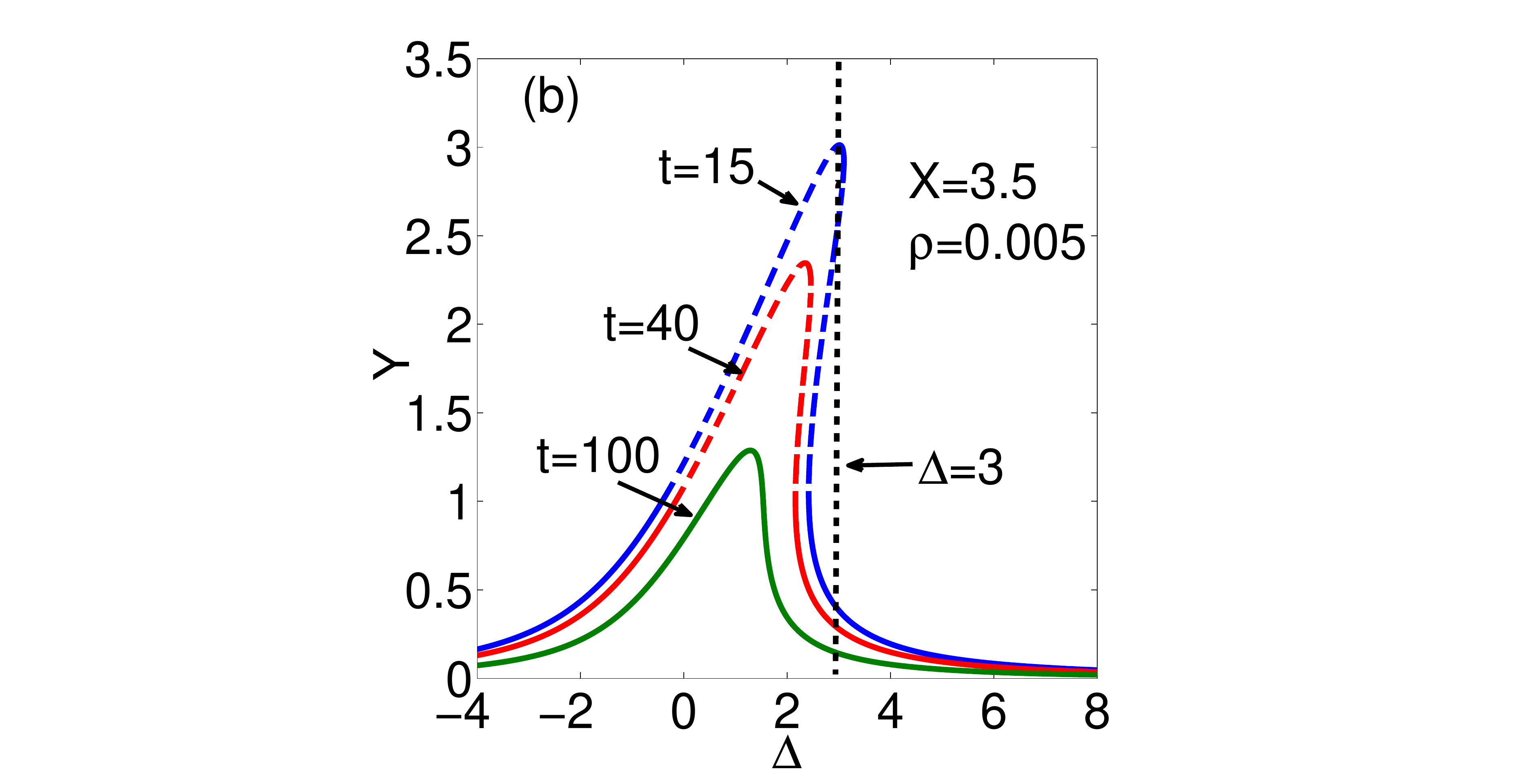,trim=3.5in 0.05in 3.8in 0.4in,clip=true, width=42mm}
\vspace{0em}
\epsfig{file=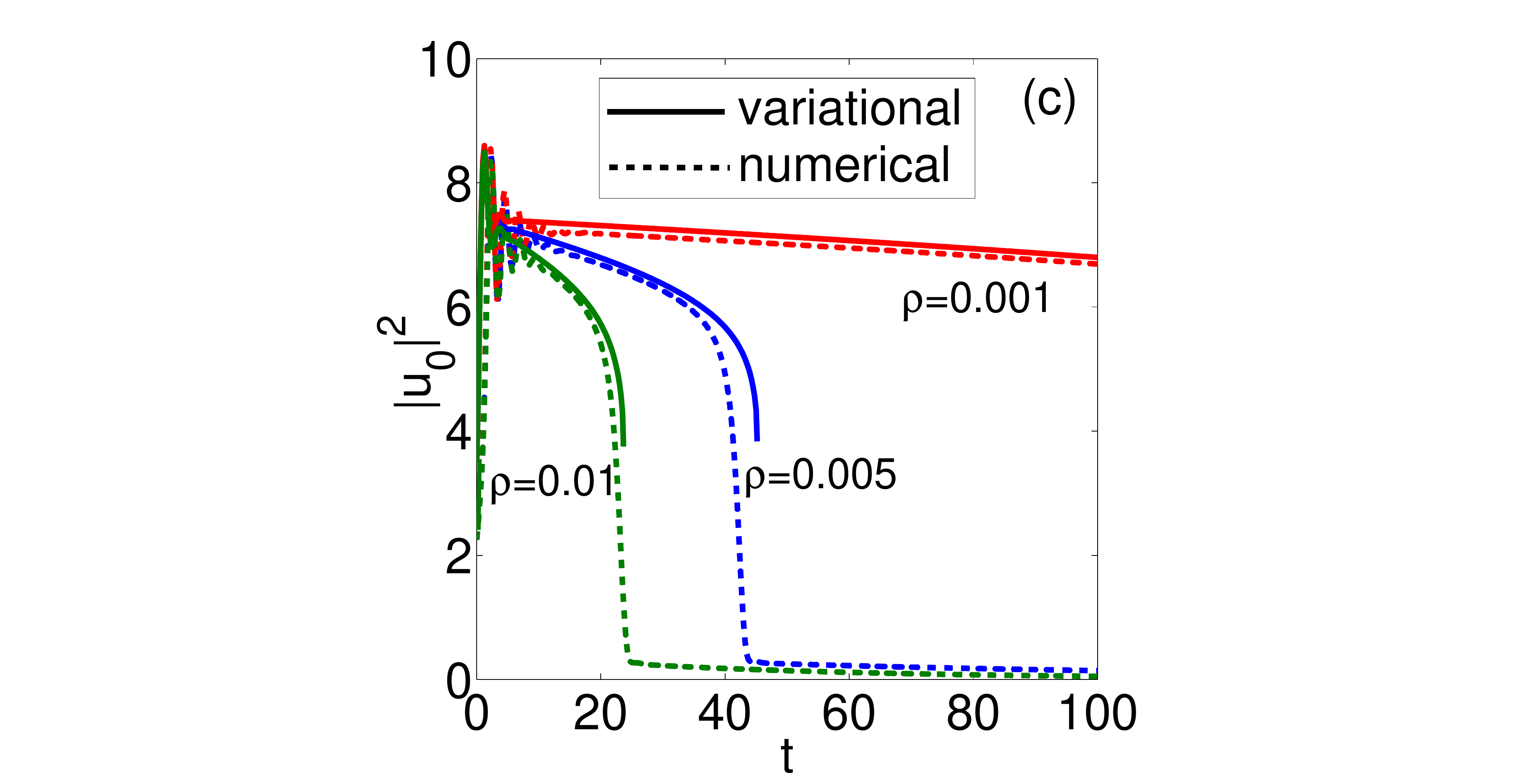,trim=3.5in 0.05in 3.8in 0.4in,clip=true, width=42mm}
\epsfig{file=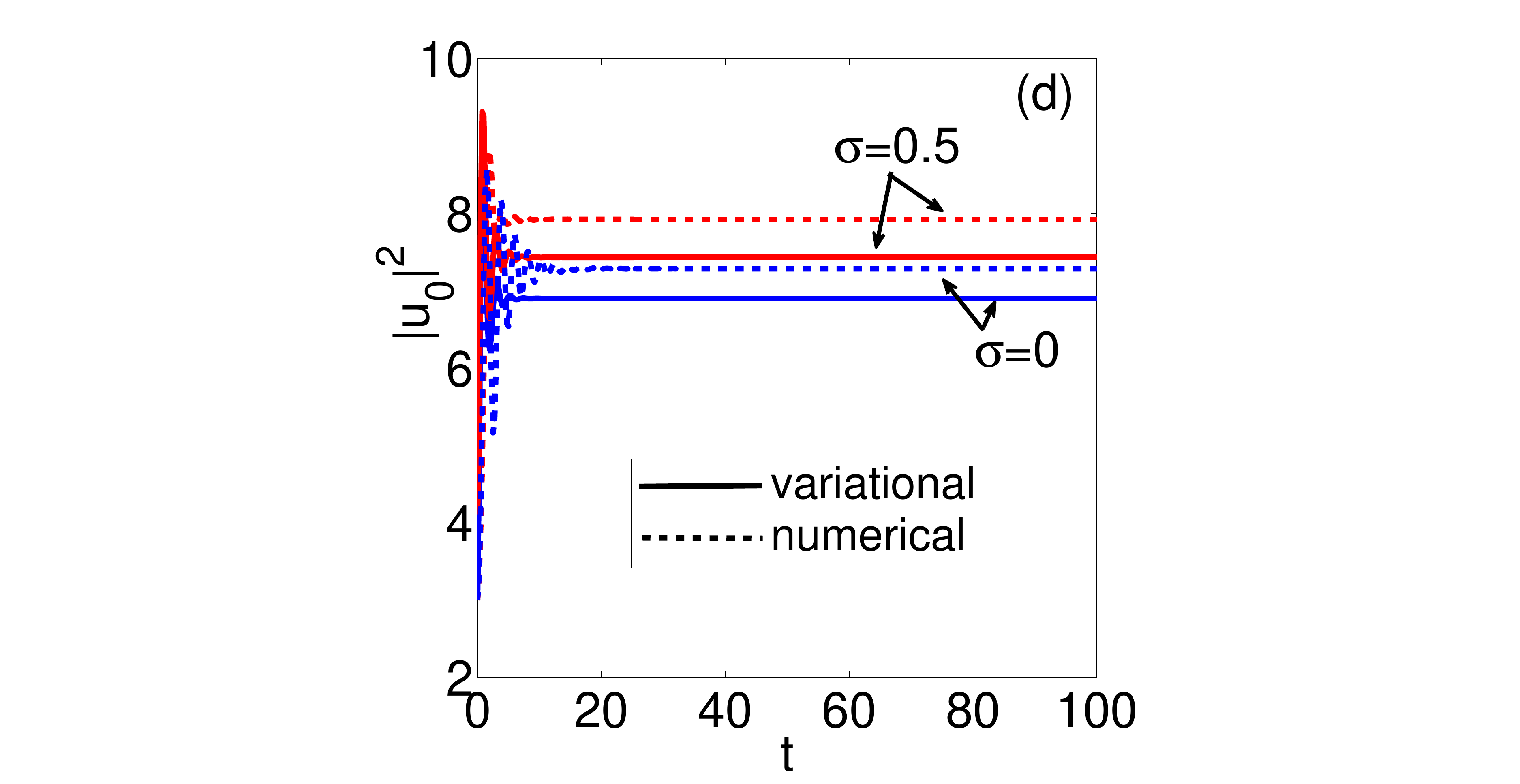,trim=3.5in 0.05in 3.75in 0.4in,clip=true, width=42mm}
\vspace{0em}
\epsfig{file=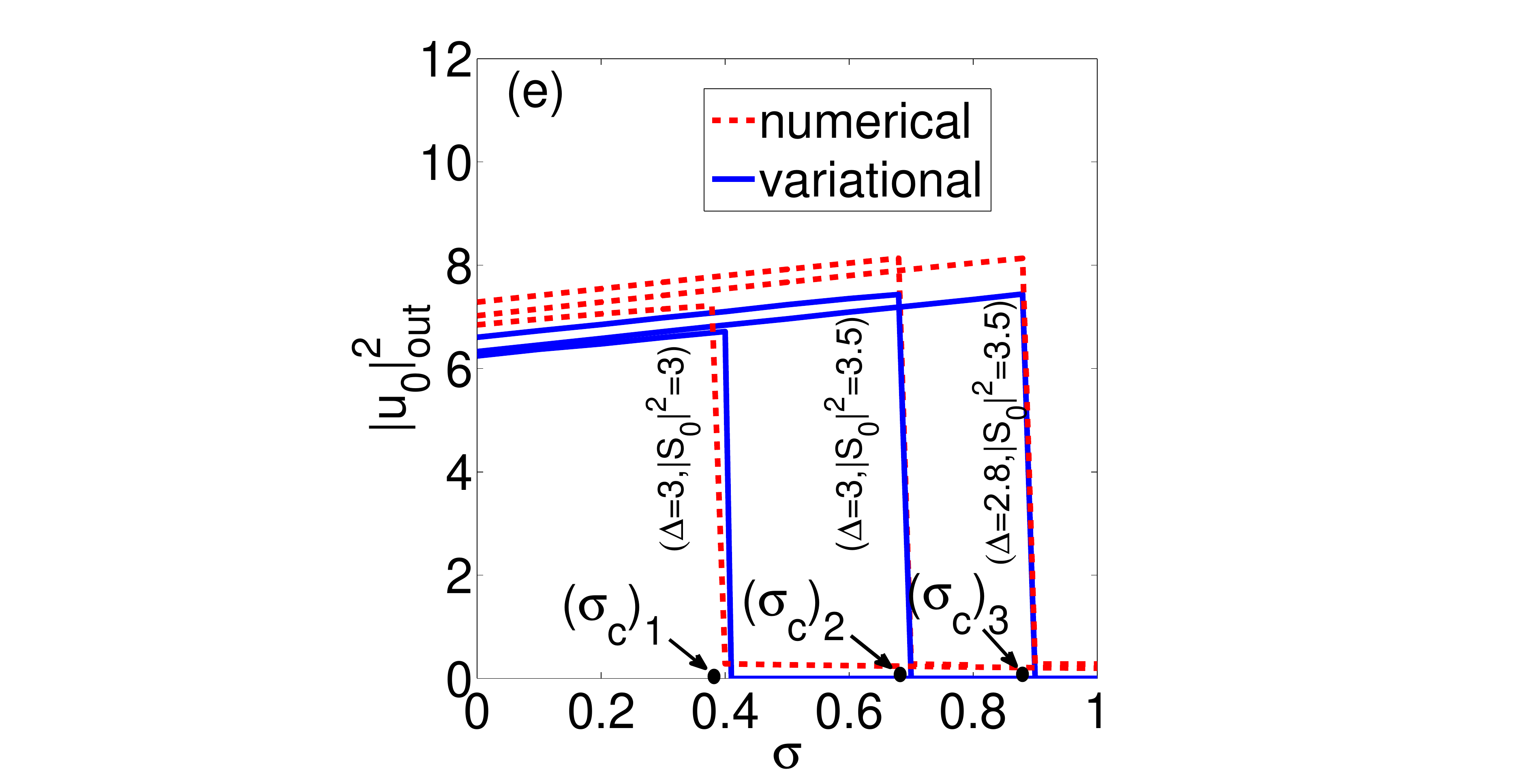,trim=3.4in 0.05in 3.8in 0.4in,clip=true, width=43mm}
\epsfig{file=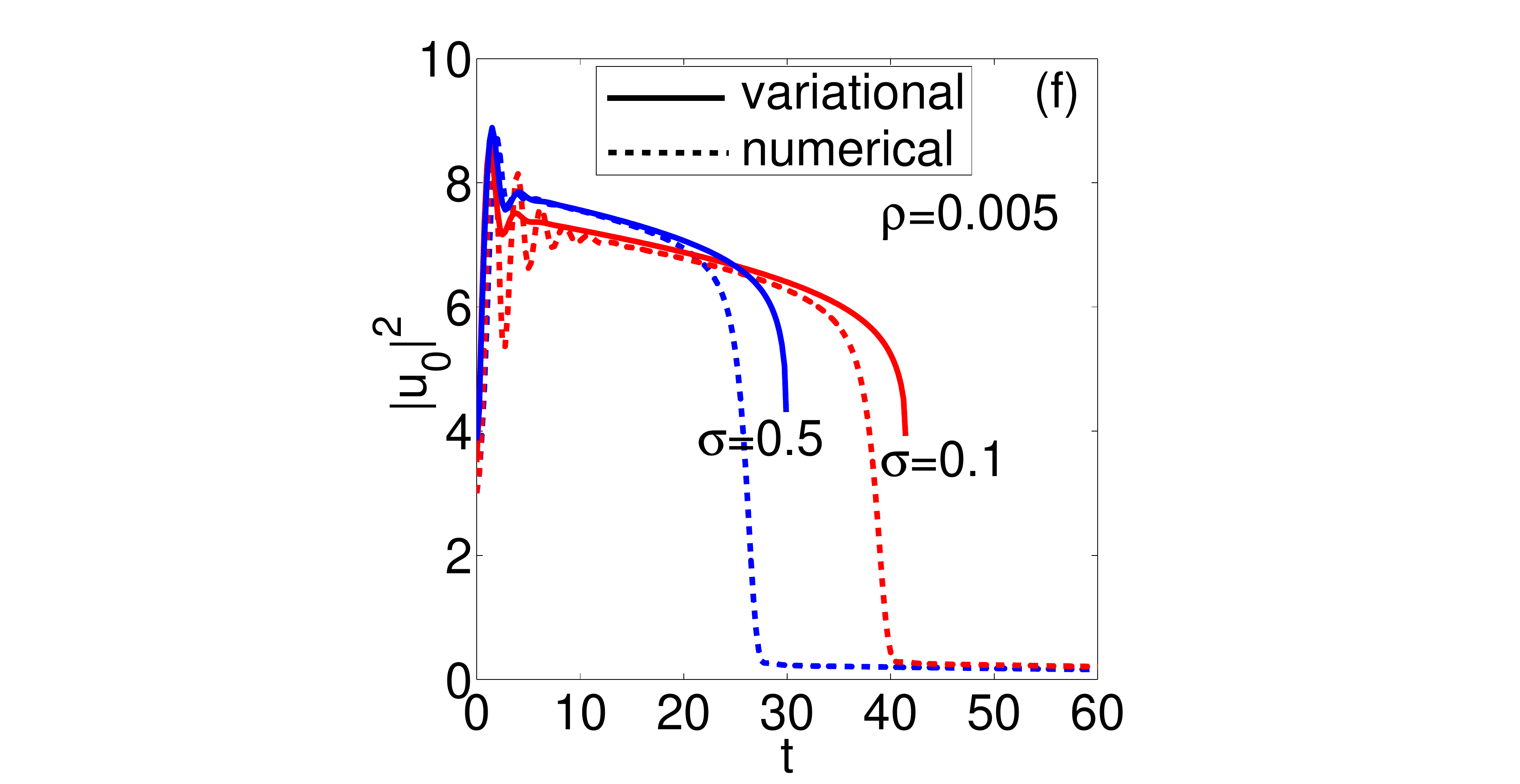,trim=3.5in 0.05in 3.8in 0.4in,clip=true, width=42mm}
\vspace{0em}
\caption{(Color online) (a) The evolution of CS over round-trip time in three dimension for $\rho=0.005$. (b) Kerr bistability in $(\Delta,Y)$ parameter space.  The variation of peak intensity $\left(|u_0|^2=E\eta/2\right)$ over the round-trip time $t$ for (c) different values of $\rho$, and for  (d) different values of $\sigma$ . (e) The variation of output peak intensity of CS with respect to $\sigma$ for three different sets of $\Delta$ and $|S_0|^2$. For each of these three cases we find three critical values of $\sigma$ [$(\sigma_c)_1\approx 0.378,~(\sigma_c)_2\approx 0.679$ and $(\sigma_c)_3\approx 0.879$] indicated by solid circles. (f) The variation of numerically simulated peak intensity over the round-trip time $t$ for different values of $\sigma$ at a fixed $\rho$.}\label{figPD}
\end{center}
\end{figure}

\noindent 
It is easy to obtain the modified steady-state CW solution of Eq.\,\eqref{LL_pd} as
\begin{equation}
X \,\exp[-2\rho t] = Y^3 - 2\Delta Y^2 +(1+ \Delta^2)Y. \label{OB_PD} 
\end{equation}  
Eq.\,\eqref{OB_PD} illustrates the steady-state CW solution of the intracavity field depends on slow time $t$. In Fig.\,\ref{figPD}(b), we plot Eq.\,\eqref{OB_PD} for $\rho=0.005$ in $(\Delta, Y)$ parameter space for a fixed $X=|S_0|^2=3.5$, where the dashed portion of the bistability curves give the unstable region. The slow-time dependence of Eq.\eqref{OB_PD} modifies the threshold values of the system parameters over $t$, as a result it also modifies the existence of CSs. It is evident from Fig.\,\ref{figPD}(a) that the CS ceases to exist beyond $t>35$ for a non-vanishing depletion coefficient ($\rho=0.005$). The impact of $\sigma$ on the dynamics of CS is not trivial and requires a detailed study. Note that, the CW bistability analysis and the MI analysis provide the same results as that of the unperturbed case. We perform the MI analysis by considering lossy ($\rho\neq0$) phase-modulated ($\sigma\neq0$) driving field as perturbations, but the direct consequence of these terms are not present in the expression of the MI gain. Only the $X$ that changes with slow-time $t$ as $X=|S_0|^2 e^{-2\rho t}$ changes the stability condition of the homogeneous steady-state solutions of the LLE.

\subsection{Perturbative analysis}
\noindent 
To grasp the effects of phase-modulated external pump on the formation and stability of CS, we adopt the standard variational analysis (see Appendix A), where the perturbation term contains the driving field $S(t)$ and linear loss as: $\epsilon \left( u \right)= S(t)-u$. Using the ansatz function $ u\left(t,\tau  \right)=\sqrt{\frac{E(t)\eta(t)}{2}}{{\left[ \text{sech}\left\{ \eta \left( t  \right) \tau  \right\} \right]}} 
\exp\left[ i \phi \left( t  \right) \right]$, we obtain a set of two coupled ordinary differential equations and one self-consistent equation describing the dynamics of a perturbed CS under pump phase-modulation as
\begin{align} 
\frac{dE}{dt}=&-2E +2S_0\, \exp(-\rho\,t) \left(\frac{E}{2\eta}\right)^{1/2} \pi\, \cos(\phi-\sigma t), \label{eqPD1}
\end{align}
\begin{align}
\frac{d{{\phi }_{p}}}{dt }=&\frac{1}{3}\eta(E-\eta) -\Delta -S_0\,\exp(-\rho\,t)\nonumber\\ &\hspace{2.0cm}\times\left(\frac{1}{2E\eta}\right)^{1/2}\pi \,\sin(\phi-\sigma t), \label{eqPD2}\\
\eta=\frac{E}{4}& +\frac{3S_0}{2\eta}\, \exp(-\rho\,t)\left(\frac{1}{2E\eta}\right)^{1/2}\pi \, \sin(\phi-\sigma t). \label{eqPD3}
\end{align}
In Fig.\,\ref{figPD}(c), we plot the evolution of peak intensity obtained from the full numerical simulation for several values of $\rho$ and compare the results with variational data by solving the set of Eqs.\,\eqref{eqPD1}-\eqref{eqPD3}. The CS ceases to exist for a critical $\rho$ which is appreciated by the sudden fall of the peak power. Note that, the maximum detuning upto which the perturbed CS can sustain is now time-dependent and can be expressed as $\Delta_{\rm {max}}(t)=\pi^2 S_0^2\, {\rm exp}(-2\rho t)/8$. The evolution of peak intensity for different values of $\sigma$ (with $\rho=0$) and the variation of peak output intensity ($|u_0|^2_{out}$) as a function of  $\sigma$ are also shown in Fig.\,\ref{figPD}(d) and (e), respectively. It is found that the steady amplitude of CS remains almost unaffected by $\sigma$ upto a critical limit ($\sigma_c$). The CS does not evolve for $\sigma>\sigma_c$. This $\sigma_c$ is not unique but depends on the set ($\Delta, |S_0|^2$). We illustrate this result in Fig.\,\ref{figPD}(e), where variational analysis accurately predicts $\sigma_c$ for three different sets of ($\Delta, |S_0|^2$) values. We further examine that, for non-zero $\rho$, $\sigma$ can limit the lifetime of CS as shown in Fig.\,\ref{figPD}(f). The power decays faster over a round trip for higher values of $\sigma$. The decay dynamics of peak power is again well predicted by variational analysis as shown by the solid lines. The variational results sustain upto the knee region of the curves because the ansatz collapses after that region.

\section{CONCLUSIONS}
\noindent 
By exploiting the standard variational technique, we study the dynamics of a perturbed CS excited inside a silicon-based microresonator where free carriers are generated owing to TPA. The pulse evolution in such a system is governed by the coupled mean-field LLE containing additional terms like TPA, IRS, and FC generation. We treat these additional terms as small perturbations and execute the variational treatment by choosing a standard $sech$ pulse as an ansatz. The variational treatment provides a set of coupled equations describing the evolution of individual pulse parameters of CS under perturbation. We suitably approximate those equations to obtain the closed-form expressions for steady amplitude, temporal and frequency shifts, etc. These closed-form analytical expressions are useful in understanding the underlying physics of complex CS dynamics under perturbations. For each perturbation, we reformulate the stability condition of the homogeneous steady-state solution of the LLE and derive explicit expressions of the threshold values of the parameters for which the bistability initiates. We perform the MI analysis to obtain the stability condition of steady-state solutions of LLE against perturbations. Exploiting the variational treatment, mathematical expressions of maximum detuning, width, and amplitude of the stable CS are derived. We solve the LLE numerically using the split-step Fourier method and demonstrate that the variational results agree well with full simulations. Finally, we consider the phase-modulated pump-field and investigate its effect on CS dynamics. We find that, for a given set of detuning and peak pump power, we may have a critical value of the phase of the pump beyond which no CS exists. In summary, our semi-analytical treatment based on the variational method provides significant insights in understanding the complex dynamics of dissipative CSs under various perturbations.

\section*{Acknowledgements}
\noindent The authors would like to thank Prof. G. P. Agrawal for useful discussions. A.S. acknowledges MHRD, India for a research fellowship.

\appendix
\section*{Appendix A: Variational method}
\label{AppendixA}
\setcounter{equation}{0}
\renewcommand{\theequation}{A{\arabic{equation}}}
\noindent 
In this appendix, the set of equations Eqs.\,\eqref{var6}-\eqref{var10} are derived using variational method. For this we write Eq.~\eqref{LL} in the form of a perturbed nonlinear Schr\"{o}dinger equation~\cite{Anderson, GPA,Vahala}:
\begin{equation} \label{nls1}
    i\frac{\partial u}{\partial t} -a\frac{\partial^{2}u}
    {\partial\tau^{2}}+ |u|^{2}u -\Delta\,u = i\epsilon(u),
\end{equation}
where $a= sgn(\beta_2)$ and $\epsilon(u)$ contains all the perturbation terms including the external gain and the loss of the system:
\begin{align} \label{nls2}
    \epsilon(u) = S -u -K|u|^2u & -i\tau_R u  \frac{\partial{|u|}^{2}}{\partial\tau} \nonumber \\ &\hspace{-1.3cm} -\tau_{sh}\frac{\partial ({|u |}^{2} u)}{\partial \tau}  -\left(\frac{1}{2}+i\mu\right)\phi_c u.
\end{align}
Next, we follow a standard procedure \cite{GPA} and introduce the appropriate Lagrangian density ($\cal{L_D}$) for Eq.\ \eqref{nls1} as
\begin{align} \label{nls3}
{\cal L_D}= \frac{i}{2} \left( u^* \frac{\partial u}{\partial t} - u \frac{\partial u^*}{\partial t} \right) & +a\left|\frac{\partial u}{\partial \tau}\right|^2 + \frac{1}{2}|u|^4 -\Delta\,|u|^2 \nonumber \\ 
 &~~~~~~~~+i\left( \epsilon^*\,u -\epsilon \, u^*  \right)
\end{align}
 and integrate over $\tau$ using the ansatz in Eq.\ \eqref{nls0} to obtain the reduced Lagrangian ($L=\int_{-\infty}^\infty  {\cal{L_D}}\,d\tau $) as
\begin{align} \label{nls4}
    L =& -E\left( \frac{\partial \phi }{\partial t }+ \Omega_p \frac{\partial {{\tau }_{p}}}{\partial t } \right) + \frac{\eta E^2}{6} +a\left( \frac{1}{3}E\eta^2 +E \Omega_p^2 \right) \nonumber \\ 
&~~~~~~~~~~~~~~~~~~ -\Delta\, E +i\int_{-\infty }^{\infty}\left( \epsilon^*\,u -\epsilon \, u^*  \right)\,d\tau.
\end{align}
We use the Euler-Lagrange equation for each pulse parameter to obtain a set of coupled ODEs for the five parameters that describe the overall soliton dynamics \cite{GPA,Hasegawa-K}. These equations govern the evolution of pulse energy $E$, temporal position $\tau_p$, frequency shift $\Omega_p$, phase $\phi$, and inverse of pulse width $\eta$ as
\begin{align}
    \frac{dE}{dt }&=2{\rm Re}\int\limits_{-\infty }^{\infty }{\epsilon {{u}^{*}}}\,d\tau, \label{var1}\\   
    \frac{d{{\tau}_{p}}}{dt }& = 2a\Omega_p +\frac{2 }{E}{\rm Re} \int\limits_{-\infty }^{\infty }{\left( \tau -{{\tau }_{p}} \right)\left( \epsilon {{u}^{*}} \right)\,d\tau }, \label{var2}\\
    \frac{d\Omega_p }{dt} &= -\frac{2\eta}{E}{\rm Im}\int\limits_{-\infty }^{\infty }{\tanh\left[ \eta \left( \tau -{{\tau }_{p}} \right) \right](\epsilon {{u}^{*}})\,d\tau }, \label{var3}\\
    \frac{d\phi}{dt} &= \frac{1}{3}E\eta -\Omega_p\frac{d{{\tau}_{p}}}{dt } -\Delta 
   +a\left( \frac{1}{3}\eta^2 +\Omega_p^2 \right) \nonumber\\
          &\hspace{3cm}+\frac{1}{E}{\rm Im}\int\limits_{-\infty }^{\infty}{\epsilon {{u}^{*}}}\,d\tau, \label{var4}\\
    \eta =& -a\frac{E}{4} - a\frac{3}{E}{\rm Im}\nonumber\\ &\times\int\limits_{-\infty }^{\infty }{\left\{\frac{1}{2\eta}-(\tau-\tau_p) \tanh\left[ \eta \left( \tau -{{\tau }_{p}} \right) \right]\right \}(\epsilon {{u}^{*}})\,d\tau },       \label{var5}
\end{align}
where, Re and Im indicate the real and imaginary parts, respectively. The ODEs Eqs.\ \eqref{var1}-\eqref{var4}, and self-consistent Eq.\ \eqref{var5} are  coupled with each other. The final step is to evaluate all the integrals using $\epsilon(u)$ given in Eq.\ \eqref{nls2}. It results in the following set of five coupled (four ODEs and one self-consistent) equations [Eqs.\,\eqref{var6}-\eqref{var10}].

\subsection*{Stationary values of CS parameters}
\subsubsection{Unperturbed case}
\noindent
The reduced variational equations [Eqs.\,\eqref{var6}-\eqref{var10}] may lead to the closed-form expression of the CS parameters that describe the steady-state.
From Eq.\,\eqref{var6}, the steady-state energy ($dE/dt=0$) of the unperturbed CS is obtained by assuming $\phi$ to be small and $\cos\phi\approx 1$ as
\begin{align} \label{A11}
E_{\rm sat}\approx\frac{\pi^2S^2}{2\eta_{\rm sat}}; ~ {\rm and}~ |u_0|^2_{\rm sat} = \frac{E_{\rm sat}\eta_{\rm sat}}{2}\approx\frac{\pi^2S^2}{4}.
\end{align}
Similarly, from Eq.\,\eqref{var10} we get the steady value of $\eta$ or the temporal pulse width $\tau_w$ by considering $\sin\phi\approx0$ as
\begin{align} \label{A12}
&\eta_{\rm sat}\left(\approx\frac{E_{\rm sat}}{4}\right)=\sqrt{\frac{|u_0|^2_{\rm sat}}{2}}; 
\nonumber \\
&\hspace{0.0cm}{\rm and}~\tau_{w\, \rm sat}\left(=2\eta_{\rm sat}^{-1} \right)=2\sqrt{2}/|u_0|_{\rm sat}.
\end{align}
In the unperturbed case, from Eqs.\,\eqref{var7} and \eqref{var8} we can show $\tau_{p\,\rm sat}=0$ and $\Omega_{p\,\rm sat}=0$, which means the CS will move without any group delay  keeping its initial frequency intact. 
Considering $d\phi/dt=0$ and $\sin\phi\approx0$ in Eq.\,\eqref{var9} and substituting the values of $E_{\rm sat}$ and $\eta_{\rm sat}$ we can obtain the well-know expression of maximum detuning $\Delta_{\rm max}$ as
\begin{align} \label{A13}
\Delta_{\rm max} =\frac{\pi^2S^2}{8}.
\end{align}
 
\subsubsection{Under perturbation}
\noindent 
{\bf i.\,\,TPA:}
In the case of TPA, $|u_0|_{\rm sat}^2$ can be calculated from Eq.\,\eqref{var6} and Eq.\,\eqref{var10} with the same assumptions that are considered in the unperturbed case. We derive Eq.\,\eqref{tpa_maxparam} from Eqs.\,\eqref{var6}, \eqref{var9}, and \eqref{var10}. Eliminating $\sin\phi$ term from Eq.\,\eqref{var9} by substituting Eq.\,\eqref{var10} and considering the steady-state ($d\phi/dt=0$ and $dE/dt=0$), we get 
\begin{align}\label{A13TPA}
\eta^2+\Delta-E\eta/2=0;~~ \cos\phi=\frac{E(1+K\eta E/3)}{\pi S(E/2\eta)^{1/2}}.
\end{align} 
Combining the expressions in Eq. \eqref{A13TPA}  for maximum critical parameter $\cos\phi=1$ and using the relation $E_{\rm sat}=4\eta_{\rm sat}$, we finally achieve Eq.\,\eqref{tpa_maxparam}.
\vspace{0.2cm}\\
\noindent {\bf ii.\,\,FC:}
The closed form expression of saturated energy $E_{sat}$ can be approximately calculated from Eqs.\,\eqref{var6} and \eqref{var10}. $\eta_{sat}$ can be approximately written from Eq.\,\eqref{var10} as $\eta_{sat}\approx (2 +\mu\theta E_{sat})E_{sat}/8$. Setting $dE_{sat}/dt=0$ we substitute $\eta_{sat}$ in the Eq.\,\eqref{var6}. Assuming the arguments of $sech$ and $cos$ to be small, we may have  ${\rm sech}{\left({\pi \Omega_p^{sat}}/{2 \eta_{sat}}  \right)}\,\cos{\phi}\approx1$. Further neglecting the higher-order terms of $E_{sat}$ associated with $\theta$ (which is small), we finally arrive 
 \begin{align} \label{Afc_energy}
E_{sat}\approx \sqrt{2}\,\pi S/\left(1+ \pi \mu \theta S/2\sqrt{2}\right).
\end{align}
From Eq.\,\eqref{var8} we can obtain the the closed-form expression of the saturated frequency blueshift $\Omega_{p\,\rm FC}^{sat}$ by setting $d\Omega_{p}^{sat}/dt=0$. Now considering above assumptions, we can express the saturated frequency as
\begin{align}\label{Afc_freq}
\Omega_{p_{\,\rm {FC}}}^{sat}\approx\sqrt{2}\,\mu\theta (E_{sat}\,\eta_{sat})^{5/2}/15\pi S.
\end{align}
The analytical expression of temporal shift can be obtained by substituting Eq.\,\eqref{Afc_freq} into the Eq.\,\eqref{var7} as
\begin{align} \label{Afc_taup}
\tau_p(t) \approx\left(-2\,\Omega_{p_{\,\rm {FC}}}^{sat}-\frac{7}{72}\theta E_{sat}^2\right)t.
\end{align}

Like TPA, we derive $\Delta_{\max}^{\rm FC}$ [see Eq. \eqref{FC_maxparam}] assuming the frequency shift $\Omega_{p_{\,\rm {FC}}}^{sat}$ to be small which we verified using variational expression. For the given parameters ($X=3.5$, $\Delta=3$) the value turns out to be $\Omega_{p_{\,\rm {FC}}}^{sat}=0.08$ which supports our assumption.

\section*{Appendix B: Intracavity modulation instability}
\label{AppendixB}
\setcounter{equation}{0}
\renewcommand{\theequation}{b{\arabic{equation}}}
\noindent 
In this appendix, we study the intracavity MI, where time-stationary periodic structures are generated from the breaking of a homogeneous wave \cite{grelu}. This linear stability analysis of the steady-state CW solutions includes the perturbations TPA, FC, IRS, and lossy phase-modulated driving field. The analysis is performed by introducing the following ansatz to the Eq.\,\eqref{LL}  \cite{Haelterman,TH-DM}
\begin{align} \label{MI_ansatz}
u(t,\tau)= u_s +a_+(t)e^{i\Omega\,\tau}+ a_-(t)e^{-i\Omega\,\tau},
\end{align} 
where $u_s$ is the steady-state CW solution of Eq.\,\eqref{LL}, $a_+$ and $a_-$ are small sideband amplitudes, and $\Omega$ is normalized sideband frequency. The substitution of Eq.\,\eqref{MI_ansatz} into Eq.\,\eqref{LL} results two coupled differential equations of the sideband amplitudes.  Linearizing the coupled differential equation with respect to $a_+$ and $a_-$ we get
\begin{align} \label{MI_matrix}
\frac{\partial}{\partial t}
\begin{bmatrix}
a_+ \\
a_-^*
\end{bmatrix}
&=
\begin{bmatrix}
a_1+a_2+ia_3 & (-K+a_4)u_s^2 \\
(-K-a_4)u_s^{*2} & a_1-a_2-ia_3
\end{bmatrix}
\begin{bmatrix}
a_+ \\
a_-^*
\end{bmatrix} \nonumber \\ &= \mathcal{M} \begin{bmatrix}
a_+ \\
a_-^*
\end{bmatrix},
\end{align} 
where $\mathcal{M}$ represents a $2 \times 2$ matrix with $a_1= (-1-2KY-\phi_c/2)$, $a_2=\tau_R Y\Omega$, $a_3=(2Y-\Delta+\delta_2\Omega^2-\mu\phi_c)$, and $a_4=(\tau_R\Omega+i)$.
In this analysis, we assume that the average carrier density cannot vary with fast-time $\tau$ and can be considered as a constant that varies with slow-time only \cite{Hansson,Hamerly,Halder}. This is justified because the spatially accumulated free carriers are governed by the carrier rate equation that follows a round-trip boundary condition \cite{Lau}. Equation\,\eqref{MI_matrix} is a system of ordinary differential equations containing the perturbation terms. The eigenvalues of the matrix  $\mathcal{M}$ represent the intracavity MI gain that we use in the main text.

\end{document}